\documentclass{nature}[a4paper]
\usepackage{bm}
\usepackage{authblk}
\usepackage{amsmath}
\usepackage{graphicx}
\usepackage{xcolor}
\makeatletter
\let\saved@includegraphics\includegraphics
\AtBeginDocument{\let\includegraphics\saved@includegraphics}
\renewenvironment*{figure}{\@float{figure}}{\end@float}
\makeatother
\usepackage[labelfont=bf]{caption}
\usepackage{cite}
\usepackage{soul}
\newcommand{\upcite}[1]{\textsuperscript{\scalebox{0.7}{\cite{#1}}}}

\title{Manipulating chiral-spin transport with ferroelectric polarization \footnotetext{$^\ast$  E-mail: xzchen@berkeley.edu, rramesh@berkeley.edu
\\Present and permanent address of Olle Heinonen: Seagate Technology, 7801 Computer Avenue, Bloomington MN 55435}}

\author[1,$\dagger$]{Xiaoxi Huang}
\author[1,2,$\dagger$,$\ast$]{Xianzhe Chen}
\author[3,$\dagger$]{{\color{blue}Yuhang Li}}
\author[4,$\dagger$]{John Mangeri}
\author[1]{Hongrui Zhang}
\author[5]{{\color{blue}Maya Ramesh}}
\author[6]{{\color{blue}Hossein Taghinejad}}
\author[1]{Peter Meisenheimer}
\author[1]{Lucas Caretta}
\author[2,7]{Sandhya Susarla}
\author[8,9]{Rakshit Jain}
\author[10]{Christoph Klewe}
\author[6]{Tianye Wang}
\author[1]{Rui Chen}
\author[2,11]{{\color{blue}Cheng-Hsiang Hsu}}
\author[1]{Hao Pan}
\author[12]{Jia Yin}
\author[10]{Padraic Shafer}
\author[6]{Ziqiang Qiu}
\author[13]{Davi R. Rodrigues}
\author[14]{Olle Heinonen}
\author[12]{Dilip Vasudevan}
\author[4,15]{Jorge \'{I}$\tilde{\rm{n}}$iguez}
\author[5]{{\color{blue}Darrell G. Schlom}}
\author[2,11]{Sayeef Salahuddin}
\author[1,2]{Lane W. Martin}
\author[6]{{\color{blue}James G. Analytis}}
\author[8,9]{Daniel C. Ralph}
\author[3,16]{{\color{blue}Ran Cheng}}
\author[12]{Zhi Yao}
\author[1,2,6,$\ast$]{Ramamoorthy Ramesh}

\affil[1]{Department of Materials Science and Engineering, University of California, Berkeley, CA 94720, USA}
\affil[2]{Materials Sciences Division, Lawrence Berkeley National Laboratory, CA 94720, USA}
\affil[3]{Department of Electrical and Computer Engineering, University of California, Riverside, CA 92521, USA}
\affil[4]{Materials Research and Technology Department, Luxembourg Institute of Science and Technology (LIST), Avenue des Hauts-Fourneaux 5, L-4362, Esch/Alzette, Luxembourg}
\affil[5]{Department of Materials Science and Engineering, Cornell University, Ithaca, NY 14853, USA.}
\affil[6]{Department of Physics, University of California, Berkeley, CA 94720, USA}
\affil[7]{School for Engineering of Matter, Transport and Energy, Arizona State University, Tempe, AZ 85281, USA}
\affil[8]{Department of Physics, Cornell University, Ithaca, NY 14853, USA}
\affil[9]{Kavli Institute at Cornell for Nanoscale Science, Ithaca, NY 14853, USA}
\affil[10]{Advanced Light Source, Lawrence Berkeley National Laboratory, Berkeley, CA 94720, USA}

\affil[11]{Department of Electrical Engineering and Computer Sciences, University of California, Berkeley, CA 94720, USA}
\affil[12]{Applied Mathematics and Computational Research Division, Lawrence Berkeley National Laboratory, CA 94720, USA}

\affil[13]{Department of Electrical Engineering, Politecnico di Bari, Via Edoardo Orabona, 4, 70126 Bari BA, Italy}
\affil[14]{Materials Science Division, Argonne National Laboratory, 9700 S Cass Ave, Lemont, IL 60439, USA}
\affil[15]{Department of Physics and Materials Science, University of Luxembourg, 41 Rue du Brill, Belvaux L-4422, Luxembourg}
\affil[16]{Department of Physics and Astronomy, University of California, Riverside, California 92521, USA}

\graphicspath{{figures/}}

\date{}                   
\begin{document}
	
	\maketitle
	
	\vspace{10pt}
	
	\begin{abstract}
		\textbf{A collective excitation of the spin structure in a magnetic insulator can transmit spin-angular momentum with negligible dissipation\upcite{kruglyak2010,lenk2011,khitun2010,kajiwara2010,chumak2015}. This quantum of a spin wave, introduced more than nine decades ago\upcite{bloch1930}, has always been 
		manipulated through magnetic dipoles\upcite{han2019,hamadeh2014,zhou2022,demokritov2004}, (\emph{i.e.}, time-reversal symmetry). Here, we report the experimental observation of chiral-spin transport in multiferroic BiFeO$\bf_3$, where 
		the spin transport is controlled by reversing the ferroelectric polarization (\emph{i.e.}, spatial inversion symmetry). The ferroelectrically controlled magnons produce an unprecedented ratio of up to 18\% rectification at room temperature. The spin torque that the magnons in BiFeO$\bf_3$ carry can be used to efficiently switch the magnetization of adjacent magnets, with a spin-torque efficiency being comparable to the spin Hall effect in heavy metals. Utilizing such a controllable magnon {\color{blue}{generation and transmission}} in BiFeO$\bf_3$, an all-oxide, energy-scalable logic is demonstrated composed of spin-orbit injection, detection, and magnetoelectric control. This observation opens a new chapter of multiferroic magnons and paves an alternative pathway towards low-dissipation nanoelectronics.}\\
	\end{abstract}

    Intense studies of spin current transport in magnetic insulators\upcite{kajiwara2010,wang2014,chen2018,wang2003, wang2019} have opened various avenues for research on spintronics\upcite{lebrun2018,wang2019spin,chen2019,han2020,dkabrowski2020, chen2021, zhang2022control} and their potential use in low-power applications\upcite{wang2003}. This trend has also triggered renewed interest in magnon physics, which can be controlled via time-reversal-symmetry design (\emph{e.g.},
    magnetic domain walls\upcite{han2019}, spin Hall currents\upcite{hamadeh2014, zhou2022},  static magnetic fields\upcite{demokritov2004}). On the other hand, recent works have shown that broken inversion symmetry and the resulting  Dzyaloshinskii-Moriya interaction (DMI) {\color{blue} have profound influence on} magnon transport in non-centrosymmetric magnets\upcite{moon2013, iguchi2015, gitgeatpong2017}. In this spirit, multiferroics are potentially of great interest, since they display broken inversion symmetry and a ferroelectric 
    polarization that can be switched by the application of an electric field, which in turn allows for control of the magnetism and, potentially, the magnons. The idea that multiferroics could be promising candidates for magnonic manipulation has been considered for some time\upcite{wang2003,kimura2003}, but despite this early interest, observation and manipulation of magnon transport in multiferroics - that is, the capability to manipulate spin transport via ferroelectric polarization reversal and resultant DMI switching- remain elusive.
    
   Here, we demonstrate non-volatile, bistable spin transport in multiferroic BiFeO$_3$ upon reversing the ferroelectric polarization. BiFeO$_3$ is a model multiferroic that exhibits multiple order parameters (\emph{i.e.}, antiferromagnetism and ferroelectricity) and intrinsic magnetoelectric coupling of these coexisting order parameters\upcite{Chu2008, Zhao2006}. In the bulk, BiFeO$_3$ possesses a rhombohedrally distorted perovskite structure in which a spin cycloid \upcite{rovillain2010}is formed with a period length of $\sim$ 64 nm \upcite{gross2017real,sando2013} and a N\'{e}el temperature of $\sim$ 643 K \upcite{Fischer1980}. The DMI in BiFeO$_3$ thin films can influence the spin cycloid  order\upcite{haykal2020, Ederer2005, wang2003}. At the same time, BiFeO$_3$ thin films possess robust ferroelectricity, with a Curie temperature $\sim$ 1100 K and a large polarization $>$ 90 $\mu$ C/cm$^2$  \upcite{Neaton2005}. In turn, BiFeO$_3$ exhibits outstanding magnetoelectric properties which have made it the focus of numerous studies hoping to utilize it as a platform to achieve low-energy (low-voltage) memory devices (\emph{e.g.}, deterministic switching of adjacent ferromagnets\upcite{Heron2014}). Specifically, achieving control over changing the magnetization states through ferroelectric polarization switching has always been a keen pursuit in these studies. 

\begin{figure}[t!]
    	\centering
    	\includegraphics[width=0.92\textwidth]{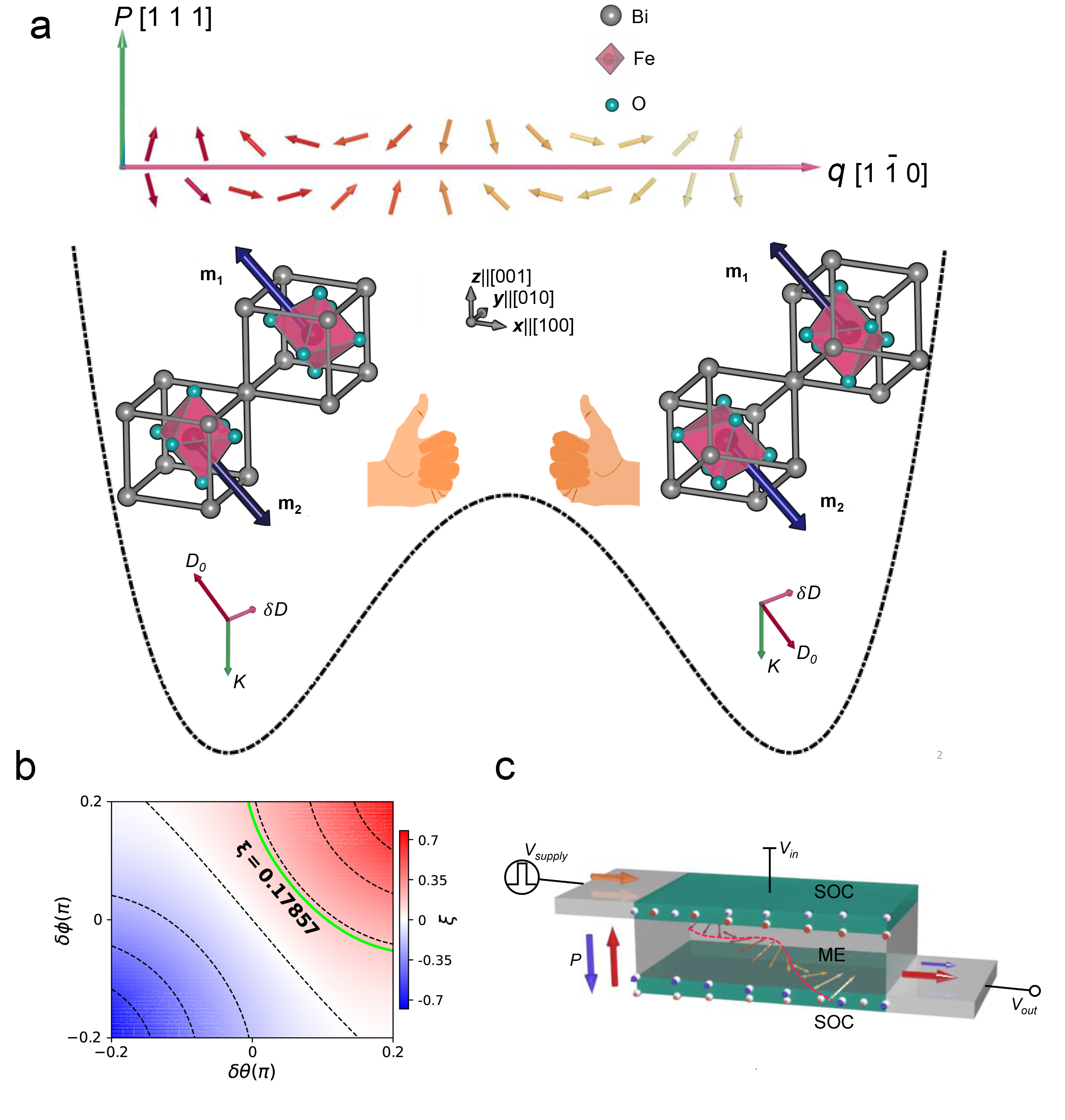}
    	\caption{\textbf{Concept of chiral magnon transport in multiferroics. 
        a, } {\color{blue}{A sketch of spin cycloid that propagates along $\bm{q}=[1 \Bar{1} 0]$. $\bm{P}$ is the ferroelectric polarization pointing along $[1 1 1]$, $\bm{D}_0$ (along $[\bar{1}\bar{1}2]$) is the intrinsic DMI vector responsible for the spin cycloid formation, and $\delta\bm{D}$ is the extrinsic DMI vector induced by the built-in electric field in BiFeO$_3$.}} 
        {\color{blue}{The double well is a schematic of the bi-stable states of ferroelectric polarization. Upon the reversal of $\bm{P}$, $\bm{D}_0$ switches by $180^\circ$ while $\delta\bm{D}$ remains unchanged, flipping the sign of the hybrid product $(\bm{D}_0 \times \delta\bm{D})\cdot\bm{K}$ with $\bm{K}$ the magnon wave vector, hence a chirality flip of the three vectors.}} 
    	\textbf{b, } {\color{blue}{Rectification ratio of spin current [defined by Eq.~\eqref{eq:rectificationratio}] as a function of $\delta{\theta}$ and $\delta{\phi}$, which are the spherical angles of $\delta\bm{D}$ relative to the direction of $\bm{D}_0$. The lime solid line marks the experimental value $\xi=0.17857$.}} 
    	\textbf{c, }Buffer logic consists of spin injection in the top SOC layer (dark green), magnetoelectric control of spin transmission in the intermediate ME layer (translucent), and spin detection in the bottom SOC layer (dark green). The switching of the polar order is controlled by $\bm{V}_{in}$. $\bm{V}_{supply}$ serves as the source for spin Hall current injection. The small (purple) and large (red) output voltages $\bm{V}_{out}$ in \textbf{c} correspond to spin transport of downward and upward ferroelectric polarization in \textbf{a} respectively.} 
    	\label{fig:1}
    \end{figure}

   {\color{blue} As illustrated in Fig.~\ref{fig:1}a, a spin cycloid propagating along $\bm{q}=[1 \bar{1} 0]$ is revealed by Nitrogen-Vacancy diamond magnetometry measurements ({{\color{blue} Extended Data Fig. 1}}), which is in good agreement with the pioneering work of Gross et al\upcite{gross2017real, sando2013}. The formation of spin cycloid has been explained in terms of a Lifshitz-like invariant\upcite{Rahmedov2012} which, in the continuum limit, amounts to an effective DMI vector $\bm{D}_0$ perpendicular to both the ferroelectric polarization $\bm{P}$ (along $[111]$) and the spin cycloid propagating direction $\bm{q}$. In BiFeO$_3$, there is another intrinsic DMI characterized by a global DMI vector parallel to $\bm{P}$, which is responsible for the small magnetization $\bm{m}$ everywhere perpendicular to the N\'eel vector $\bm{L}$. Previous studies have confirmed that the latter mechanism is much weaker so that the spin texture is dominated by $\bm{D_0}$\upcite{Rahmedov2012,competingDMI}. A reversal of $\bm{P}$ is accompanied by the reversal of the octahedral antiphase tilts\upcite{Heron2014}, which in turn flips the direction of $\bm{D}_0$\upcite{Ederer2005,Rahmedov2012}. In addition, the asymmetric gates, the existence of ferroelectric domain walls\upcite{yang2010}, and other asymmetries of external origins generate a built-in potential that induces an additional DMI vector $\delta{\bm{D}}$, which remains the same under the reversal of $\bm{P}$. Consequently, the total DMI vector $\bm{D}_0+\delta\bm{D}$ undergoes a non-$180^\circ$ switching, leading to a slight rotation of the propagation direction $\bm{q}$. To intuitively understand the ferroelectric control of magnon spin current, we define the hybrid product $(\bm{D}_0 \times \delta\bm{D}) \cdot \bm{K}$ where $\bm{K}$ is the magnon wave vector (in the thickness direction). It is well-known that the nonreciprocal propagation is allowed not only in magnons but also electrons, electromagnetic waves when they go through materials such as multiferroic materials that host broken time reversal and space inversion symmetry simultaneously. As illustrated in Fig.~\ref{fig:1}a, reversing $\bm{P}$ necessarily flips the sign of $(\bm{D}_0 \times \delta\bm{D}) \cdot \bm{K}$, hence the chirality of these three vectors. This hybrid product can also reflect a \textit{directional} non-reciprocity where the wave vector $\bm{K}$ flips sign\upcite{wang2020, moon2013,iguchi2015, gitgeatpong2017} while the DMI is kept unchanged. In other words, this single quantity captures the common feature of both directional non-reciprocity and ferroelectric modulation of the magnon spin current.
   
   Because the thickness of BiFeO$_3$ is much smaller than the spacial period ($\sim$64 nm) of the spin cycloid, we can ignore any spin texture or magnetic inhomogeneity along the magnon wave vector $\bm{K}$. Then, the transmitted spin current can be phenomenologically expressed as\cite{lebrun2018}
   \begin{align}
        J_s = \frac{G_L}{\mathcal{S}}\int d{\bf{r}}^2[\bm{L}(\bm{r})\cdot\bm{s}]^2+\frac{G_m}{\mathcal{S}}\int d{\bf{r}}^2[\bm{m}(\bm{r})\cdot\bm{s}]^2, \label{eq:rectificationratio}
   \end{align}
   where $G_L$ ($G_m$) is the phenomenological spin conductance of the N\'eel vector (total magnetization), $\bm{s}=\hat{y}$ is the polarization of spin injection from the top gate and $\mathcal{S}$ is the area of the film. As mentioned above, when the total DMI vector undergoes a non-$180^\circ$ switching, the spin cycloid profile functions $\bm{L}(\bm{r})$ and $\bm{m}(\bm{r})$ are modulated by $\bm{P}$ in a non-trivial way. As calculated in the Supplemental Materials, the resulting spin current $J_s$ can be substantially different before and after the reversal of $\bm{P}$. To quantify this difference, we define the rectification ratio as
   \begin{align}
    \xi = \frac{J_s(-\bm{P})-J_s(\bm{P})}{J_s(-\bm{P})+J_s(\bm{P})}
   \end{align}
   and plot it against the orientation of the extrinsic $\delta\bm{D}$ relative to the intrinsic $\bm{D}_0$ (along $[\bar{1}\bar{1}2]$) in Fig.~\ref{fig:1}b, where we have marked the experimental value (to be explained below) $\xi=0.17857$.


}

    Based on the predicted potential for chiral-magnon control in BiFeO$_3$, a demonstration of a scalable magnetoelectric spin-orbit logic was investigated. A schematic of this idea is illustrated in Fig.~\ref{fig:1}d.  An electric field is set up in the magnetoelectric capacitor in the $-z$ direction, resulting in the vertical component of the ferroelectricity switching from the $z$ to the $-z$ direction. To read the polarization state, a supply voltage is applied into the top spin-orbit coupled (SOC) channel (\emph{e.g.}, SrIrO$_3$ which has a relatively large spin Hall angle\upcite{huang2021novel}) in the $+x$ direction, causing the excitation of magnons in the BiFeO$_3$ via the spin current from the SOC material. The readout of the state of the switch is then enabled by the inverse spin Hall effect using SrIrO$_3$\upcite{huang2020}, where the current direction is along the $x$ direction. Owing to the magnon propagation with bistable rectification, the output voltage could acquire two stable states, rendering the buffer function of a logic gate. Especially, as compared with the first magneto-electric, spin-orbit (MESO) device proposal\upcite{manipatruni2019}, this design makes use of antiferromagnetic magnon to gauge the antiferromagnetism inside the multiferroic itself instead of relying on the adjacent ferromagnet coupled to the multiferroic ({{\color{blue} Extended Data Fig. 2}}). Early efforts on the elimination of the ferromagnet have been devoted to the electric-field controlled spin-orbit interaction\upcite{noel2020,varotto2021}, e.g. Rashba effect in two-dimensional electron gas systems\upcite{lesne2016,vaz2019}. 
    {\color{blue} {The elimination of the ferromagnet simplifies the readout architecture;  furthermore, the implementation of antiferromagnets can favor ultrafast operations frequency up to THz.}}

\begin{figure}[htbp]
    	\centering
    	\includegraphics[width=0.8\textwidth]{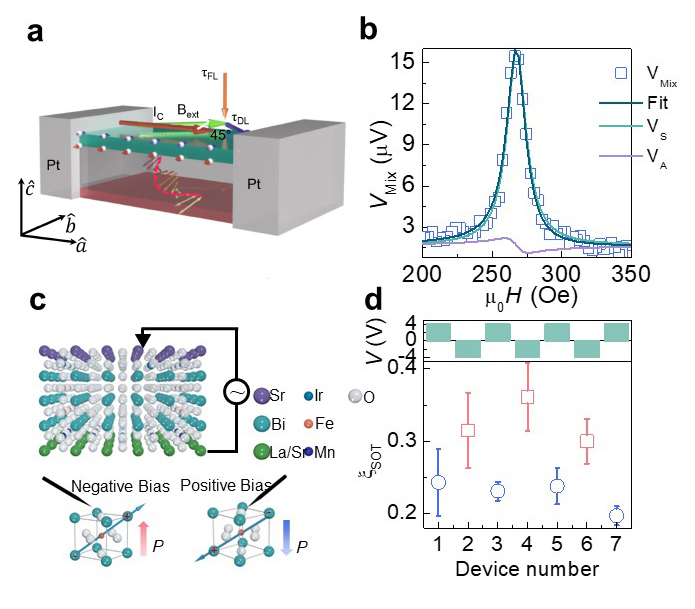}
    	\caption{\textbf{Spin transmission controlled by ferroelectric polarization in BiFeO$\bf_3$. a,} A $3$D schematic illustration for the ST-FMR measurement. Top : SrIrO$_3$, middle : BiFeO$_3$, bottom : La$_{0.7}$Sr$_{0.3}$MnO$_3$. $\bm{I}_c$ represents the radio frequency microwave current, which flows at a $45^\circ$ angle with respect to the in-plane external magnetic field represented by $\bm{B}_{ext}$. {\color{blue}{A spin cycloid that lies in $(\bar{1}\bar{1}2)$ plane, and propagates along $[1 \bar{1} 0]$ direction is dipicted by the double-headed arrows.}} \textbf{b,} Measured ST-FMR signal at room temperature for 12 dBm of applied microwave power at $4$ GHz for a sample {\color{blue}{La$_{0.7}$Sr$_{0.3}$MnO$_3$ ($20$ nm)/BiFeO$_3$ ($25$ nm)/SrIrO$_3$ ($10$ nm)}} of dimension $50\times 25$ $\mu$m$^2$. The lines are Lorentzian fits to the mixing voltage as described in methods (ST-FMR analysis section), showing both symmetric and anti-symmetric components. \textbf{c,} A schematic of a La$_{0.7}$Sr$_{0.3}$MnO$_3$ ($20$ nm) /BiFeO$_3$ ($25$ nm) /SrIrO$_3$ ($10$ nm) heterostructure subject to an external electric field, along with schematics of the BiFeO$_3$ ferroelectric polarization under negative and positive biases. \textbf{d,} The SOT efficiency measured with the ST-FMR technique as a function of bias voltages applied beforehand for samples of La$_{0.7}$Sr$_{0.3}$MnO$_3$ ($20$ nm) /BiFeO$_3$ ($25$ nm) /SrIrO$_3$ ($10$ nm) The open squares (circles) represent the SOT efficiency for the tri-layer with negative (positive) bias voltages applied on the top electrode - SrIrO$_3$.}
    	\label{fig:2}
    \end{figure}

    To experimentally explore the controllable spin transmission enabled by the ferroelectric switch in BiFeO$_3$, we first characterized the strength of spin transmission via magnons in the heterostructures of the form La$_{0.7}$Sr$_{0.3}$MnO$_3$/BiFeO$_3$ /SrIrO$_3$ (Methods, {{\color{blue} Extended Data Fig. 3,4}}). 
    We performed the measurement using a well-established technique namely spin-torque ferromagnetic resonance (ST-FMR)\upcite{huang2021novel, liu2012}(Fig.~\ref{fig:2}a), launching the spin current using the large spin Hall effect of SrIrO$_3$ and detecting the spin torque applied to the magnetic La$_{0.7}$Sr$_{0.3}$MnO$_3$ layer\upcite{merbouche2021}.  
    An example ST-FMR spectrum, composed of both symmetric and anti-symmetric components, is shown (Fig.~\ref{fig:2}b) and indicates the successful detection of the spin currents carried by magnons through the BiFeO$_3$.
    To manipulate the orientation of the ferroelectric polarization, an external electric field was applied along the out-of-plane $[0 0 1]$ to set the ferroelectric polarization of the BiFeO$_3$ as illustrated (top panel, Fig.~\ref{fig:2}c, in which the top SrIrO$_3$ and bottom La$_{0.7}$Sr$_{0.3}$MnO$_3$ layers serve as the two electrodes sandwiching the ferroelectric). Specifically, upward- (downward-) pointing polarization ($180^\circ$ switching) can be achieved by applying negative (positive) voltage larger than the coercivity of the BiFeO$_3$ on the top SrIrO$_3$ layer (bottom panel, Fig.~\ref{fig:2}c)\upcite{huang2020}. After poling the BiFeO$_3$ in each individual device to the desired polarization state, we then deposited platinum contacts onto the devices (contacting both the SrIrO$_3$ and La$_{0.7}$Sr$_{0.3}$MnO$_3$ layers) and performed ST-FMR measurements on the poled devices. By adopting an established analysis method (Methods), the SOT efficiency for the damping-like (DL) torque (symmetric component) is evaluated and summarized (Fig.~\ref{fig:2}d). We observe that the SOT efficiency can be set at a high value by negative (upward polarization) or at a low value by positive (downward polarization) voltages, indicating the presence of bistable high/low spin transport in BiFeO$_3$. This polarization-dependent spin transmission of BiFeO$_3$ is also verified by interfacial engineering ({{\color{blue} Extended Data Fig. 5}}), time-resolved X-ray ferromagnetic resonance ({{\color{blue} Extended Data Fig. 6}}) and FMR measurements ({{\color{blue} Extended Data Fig. 7, 8}}). {\color{blue}{The blocking temperature of exchange bias between La$_{0.7}$Sr$_{0.3}$MnO$_3$ and BiFeO$_3$ is well below room temperature\upcite{yi2019}, and the absence of exchange bias is also confirmed by our magneto-optic Kerr effect measurements ({{\color{blue} Extended Data Fig. 9}}), ruling out the possible contribution from exchange bias on the tuning of spin transmission.}} It is worthwhile to note that the stability of the ferroelectric polarization reveals another crucial advantage associated with spintronics produced in this manner – non-volatility. By setting the polarization direction, 
    high and low SOT efficiencies of $0.33$ and $0.23$ can be achieved, showing a high/low ratio of approximately $0.18$\%. Furthermore, the ferroelectric polarization controlled spin transmission in BiFeO$_3$ is confirmed by ST-FMR measurements on samples with {\color{blue}{opposite polarizations set by}} different interface terminations\upcite{yu2012} (Extended Data Fig. 5). This is consistent with our theoretical predictions that the polarization state influences magnon propagation {\color{blue}{and a spin transmission modulation can be achieved upon ferroelectric polarization switching}}. Further, from the aforementioned experiments it can be inferred that the spin transmission is favorable when the magnon wave vector {\color{blue}{is anti-parallel to the z-component of the polarization $\bm{P}$}}. 
    
    \begin{figure}[htbp]
    	\centering
    	\includegraphics[width=1\textwidth]{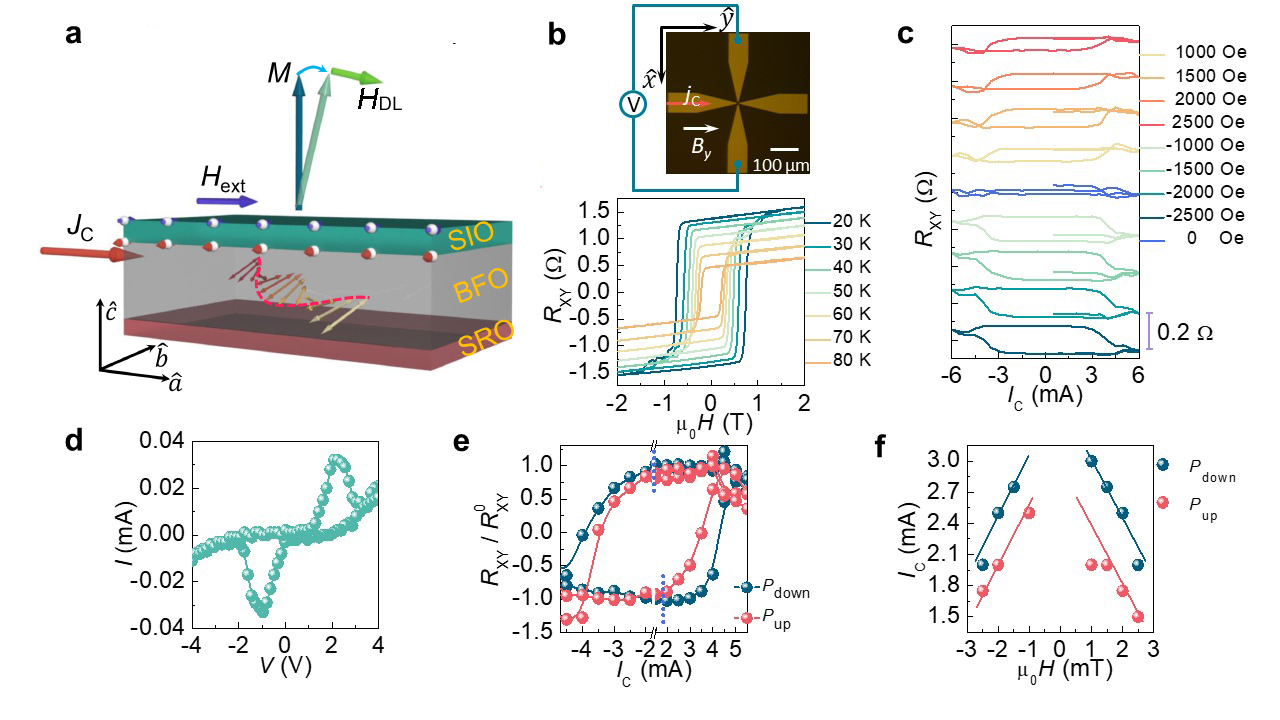}
    	\caption{\color{blue}{\textbf{Magnetization switching induced by the spin torque in BiFeO$_3$ under the control of ferroelectric polarization.  a, } A schematic demonstrating the magnetization switching of SrRuO$_3$ by the current-induced torques borne by the magnons transmitted through BiFeO$_3$ for a sample SrRuO$_3$ ($5$ nm)/BiFeO$_3$ ($25$ nm)/SrIrO$_3$ ($10$ nm). A spin cycloid that lies in $(\bar{1}\bar{1}2)$ plane, and propagates along $[1 \bar{1} 0]$ direction is depicted by the double-headed arrows. The torque carried by the transmitted magnons switches the magnetic moment (dark blue arrows) of the neighboring SrRuO$_3$ layer in the out-of-plane direction. A small assisting magnetic field ($\bm{H}_{ext}$) is applied along the pulsed current ($\bm{J}_C$) direction. \textbf{b,} An optical image of the actual Hall device used in the anomalous Hall effect (AHE) and magnetization switching measurements. Anomalous Hall resistance ($\bm{R}_{XY}$) as a function of the magnetic field at various temperatures for the heterostructure SrRuO$_3$ ($5$ nm)/BiFeO$_3$ ($25$ nm)/SrIrO$_3$ ($10$ nm) is shown on the bottom. A dc current of 100 $\mu$A is applied during the AHE measurements. \textbf{c,} $R_{XY}$ as a function of pulsed currents. External magnetic fields with different polarities and strengths are applied along the current pulse direction to assist the magnetization switching. After each current pulse, a dc current of 100 $\mu$A is applied to detect the change in $\bm{R}_{XY}$. The hysteretic magnetization switching loops are shifted with respect to the magnetization switching loop at zero assisting field for display. The current pulses have a pulse width of 1 ms. \textbf{d,} An example IV curve collected during the ferroelectric switching experiments for the heterostructure SrRuO$_3$ ($5$ nm)/BiFeO$_3$ ($25$ nm)/SrIrO$_3$ ($10$ nm). \textbf{e,} An example ferroelectric polarization controlled magnetization switching measurement. The magnetization switching behavior for SrRuO$_3$ ($5$ nm)/BiFeO$_3$ ($25$ nm)/SrIrO$_3$ ($10$ nm) heterostructure with the polarization of BiFeO$_3$ pointing from the bottom (top) electrode to the top (bottom) electrode is shown in red (blue). The $\bm{R}_{XY}$ is normalized for better display. \textbf{f,} The magnitudes of critical switching current as a function of the applied external magnetic fields are summarized. The critical switching current for SrRuO$_3$ ($5$ nm)/BiFeO$_3$ ($25$ nm)/SrIrO$_3$ ($10$ nm) heterostructure with upward (downward) ferroelectric polarization is shown in red (blue). The lines are a guide to the eye.}}
    	\label{fig:3}
    \end{figure}
    \begin{figure}[ht!]
    	\centering
    	\includegraphics[width=0.67\textwidth]{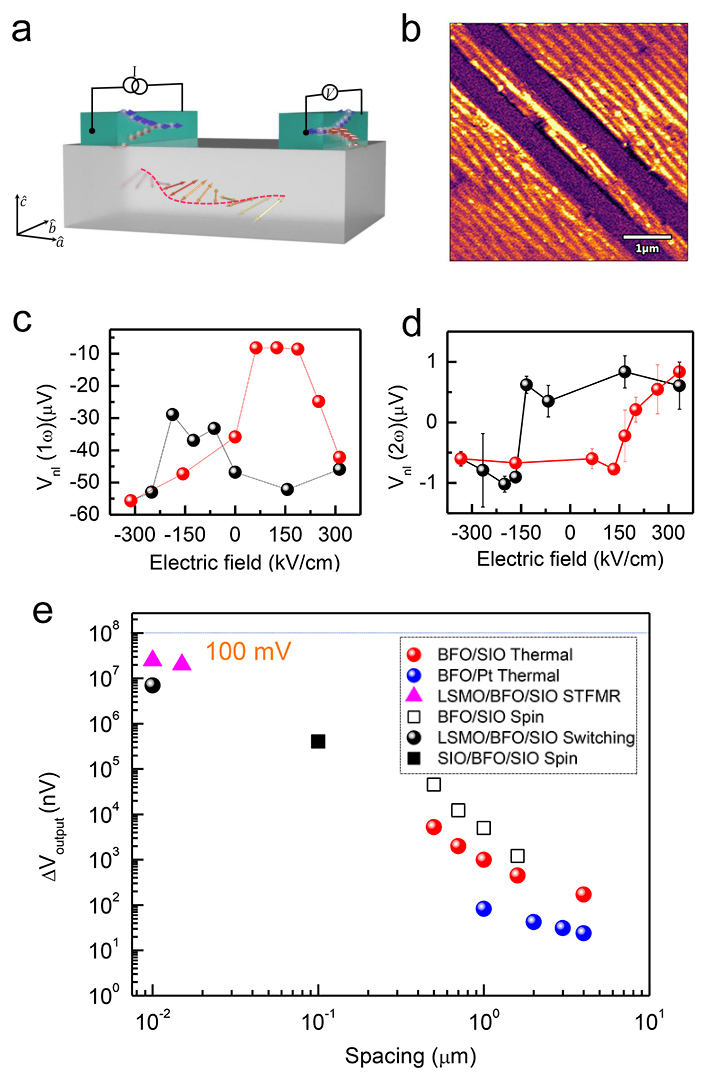}
    	\caption{\textbf{Magnetoelectric spin-orbit logic based on controllable magnon transport in BiFeO$\bf_3$. a,} Schematic of the non-local device formed with a magnetoelectric test structure.
    	\textbf{b,} Piezo-force microscopy image of the non-local device structure. The channel width is 0.5 $\mu$m.  \textbf{c,} Hysteresis loop from spin magnon readout in SrIrO$_3$ with first harmonic measurements for a device with a 500 nm channel width. \textbf{d,} Hysteresis loop from thermal magnon readout in SrIrO$_3$ with second harmonic measurements in a device with a 1 $\mu$m channel width. \textbf{e,} Output voltage scaling with channel spacing for measurements in nonlocal devices (points for spacings 0.5 $\mu$m and above), vertical structure of SIO/BFO/SIO when BFO is 100 nm, and estimates of voltages that could be obtained for vertical transport of magnons through 10 and 15 nm of BiFeO$_3$. The power is fixed at 3 mW.
    	\\}
    	\label{fig:4}
    \end{figure}

   {\color{blue}{To corroborate the presence of a magnon transmission modulation in BiFeO$_3$, magnetization switching experiments were carried out to serve as a direct demonstration of the strength of the spin torques under ferroelectric polarization modulation. Here, we investigate deterministic magnetization switching of a SrRuO$_3$ layer with perpendicular magnetic anisotropy (PMA) by the spin torque carried by the BiFeO$_3$. Current-induced magnetization switching measurements were carried out for SrRuO$_3$ ($5$ nm)/BiFeO$_3$ ($25$ nm)/SrIrO$_3$ ($10$ nm) heterostructure with BiFeO$_3$ polarized to either upward or downward orientations (Fig.~\ref{fig:3}a). The switching schematic is also demonstrated (Fig.~\ref{fig:3}a), where the magnetic moment of the ferromagnetic SrRuO$_3$ is switched between positive and negative directions consistently and reversibly. The magnetic switching was detected by the anomalous Hall effect in the SrRuO$_3$ layer, in which magnetic moments lying on the positive and negative directions can be distinguished by positive and negative values of anomalous Hall resistance. By sweeping an external magnetic field in the out-of-plane direction (Fig.~\ref{fig:3}b), a hysteretic loop is developed for anomalous Hall resistance ($\bm{R}_{XY}$), when the external magnetic field exceeds the coercivity of SrRuO$_3$(Fig.~\ref{fig:3}b). Furthermore, $\bm{R}_{XY}$ as a function of magnetic field was repeated at temperatures from $20$ K to $80$ K, confirming the high quality of our heterostructure, and no shift in the hysteretic loop with respect to zero magnetic field is observed, indicating the absence of exchange bias effect between ferromagnetic SrRuO$_3$ and antiferromagnetic BiFeO$_3$.  Then magnetization switching measurements were performed in the same device, and $\bm{R}_{XY}$ as a function of current pulses was recorded. Notably, a switching of magnetization was observed by both positive and negative currents, when the magnitudes of current pulses exceed the magnitude of critical switching current, and a saturation of $\bm{R}_{XY}$ is observed with large current pulses, excluding the possible participation of thermal effects in the switching measurements. Moreover, no magnetization switching is observed at zero assisting magnetic field and the switching chirality can be changed by the reversal of polarity of assisting magnetic fields, ruling out the possibility that the magnetization switching is caused by current-induced Oersted field. To further corroborate our observation of ferroelectric polarization-controlled spin transport with ST-FMR, ferroelectric polarization-dependent magnetization switching measurements were carried out. To start with, the polarization of BiFeO$_3$ was switched to either an upward or downward orientation by an external electric field. An example IV curve during the ferroelectric polarization switching measurements is displayed in Fig.~\ref{fig:3}d, in which the switching of ferroelectric polarization is associated with current peaks in Fig.~\ref{fig:3}d. Then, the same magnetization switching measurements were conducted on devices with distinct ferroelectric polarization states. A typical switching loop for heterostructures with upward (downward) polarization is displayed in red (blue) solid spheres in Fig.~\ref{fig:3}e. Strikingly, distinct critical switching currents for upward and downward polarization were observed in both positive and negative current directions. The critical switching currents at different assisting magnetic fields for upward and downward polarization are summarized in Fig.~\ref{fig:3}f, where upward polarization shows a smaller switching current than downward polarization statistically. Therefore, two conclusions can be drawn based on our experimental observations. First, BiFeO$_3$ has the capability to carry spin information via antiferromagnet magnons, and the spin torque arising from BiFeO$_3$ is sufficiently efficient to switch the magnetization of SrRuO$_3$. More importantly, the spin transmission in BiFeO$_3$ is controlled by ferroelectric polarization, and the smaller critical switching current density for the upward polarization corresponds to a higher spin transmission, in good agreement with the ferroelectric polarization controlled-SOT efficiency study with the ST-FMR technique. The current-induced magnetization switching measurements reinforce the idea that the upward polarization is favorable for magnon transmission.}}

   We further investigated the experimental demonstration of MESO via electrically controllable spin transport {\color{blue} using a lateral nonlocal measurement.} 
   A schematic of the nonlocal measurement is provided (Fig.~\ref{fig:4}a); it consists of spin-orbit input and output via parallel source and detector wires, as well as magnetoelectric control. Fig.~\ref{fig:4}b illustrates well-ordered $71^\circ$ ferroelectric domains, showing the high quality of the BiFeO$_3$ films. A nonlocal, quasi-static measurement was performed (Fig.~\ref{fig:4}c), varying the magnitude of the electric-field pulse applied in-plane across the channel from negative to positive and back again, and after each pulse measuring (over $100$ seconds)  non-local voltage signals due to magnon transport between the source and detector wires {\color{blue}{(Methods)}}. We observe a butterfly-shaped and hysteretic response in the magnon current launched by the spin Hall effect (first-harmonic voltage, Fig.~\ref{fig:4}c) and the magnon current launched thermally by the local spin Seebeck effect, respectively (second-harmonic voltage, Fig.~\ref{fig:4}d), in which the coercive fields match each other.
   {\color{blue} Besides, the electric field modulations of the output voltage can be obtained with dc methods, which can minimize the inductive and capacitive effect ({{\color{blue} Extended Data Fig. 10,11}}).} 
    Figure \ref{fig:4}e presents measurements of the magnitude of these spin-transport signals as a function of changing the spacing between the source and detector wires from 500 nm to 4 $\mu$m. 
   The open squares reflect the magnitude of the signals associated with spin currents launched by the spin Hall effect (first harmonic voltage) and the red circles reflect signals launched thermally by the spin Seebeck effect (second harmonic voltage).  
   Remarkably, the output voltage detected in BiFeO$_3$/SrIrO$_3$ stacks ($\sim 1$ $\mu$V) is approximately one order of magnitude larger than the counterparts in BiFeO$_3$/Pt ($\sim 0.1$ $\mu$V)\upcite{parsonnet2022}, which is  supported by the large spin-torque efficiency of SrIrO$_3$ previously observed with the ST-FMR measurement \upcite{huang2021novel, wang2019large} {\color{blue} {and the controlled nonlocal measurements of NiFe$_2$O$_4$/Pt and NiFe$_2$O$_4$/SrIrO$_3$ (Extended Data Fig. 13,14}}).
   The distance dependence for both types of signals in BiFeO$_3$ fits well to the expectation for diffusive magnon propagation\upcite{cornelissen2015}
    \begin{equation}
    	R_{\rm{nl}} = \frac{C}{\lambda}\frac{\exp(d/\lambda)}{1-\exp(2d/\lambda)}
    \end{equation}
    with effective diffusion lengths $\sim 0.17$ $\mu$m (first harmonic) and $\sim 0.25$ $\mu$m (second harmonic).
    
    Using this result and the spin-torque efficiencies obtained above {\color{blue}{(Extended Data Fig. 5)}}, we can calculate the inverse-spin-Hall output voltage that would be obtained for a vertical geometry with BiFeO$_3$ thicknesses of $10$ and $15$ nm using\upcite{shikoh2013}
    \begin{equation}
    	V_{\rm{ISHE}} = \frac{\theta_{\rm{SOT}}\lambda_{\rm{SD}}\tanh(\frac{t_{\rm{SIO}}}{2\lambda_{\rm{SD}}})L}{t_{\rm{SIO}}\sigma_{\rm{SIO}}}J_{\rm{S}}
    \end{equation}
    where we use the values $\theta_{\rm{SOT}}$ = $0.4$ is the spin-torque efficiency,  $\lambda_{\rm{SD}} = 1.4$ nm \upcite{nan2019} is the spin-diffusion length of SrIrO$_3$, $t_{\rm{SIO}} = 10$ nm is the thickness, $\sigma_{\rm{SIO}} = 1 \times 10^5$ $\Omega^{-1} m^{-1}$ is the conductivity, $L = 10$ nm is the wire length. The output voltage is on the order of $10$ mV, which is approaching the MESO target voltage of $100$ mV. A prototypical magnon-mediated MESO logic device and its circuit modeling utilizing this ferroelectrically controlled magnon propagation are shown in Extended Data Fig. 5. This observation sheds light on implementing multiferroics as the potential spin carriers and opens new possibilities of energy-efficient magnonic spin-logic devices.\\
    
    \vspace{1000pt}
    \newpage
    \noindent\textbf{References}
    \bibliographystyle{naturemag}
    \bibliography{MESO-Revised}

\begin{thebibliography}{10}
\expandafter\ifx\csname url\endcsname\relax
  \def\url#1{\texttt{#1}}\fi
\expandafter\ifx\csname urlprefix\endcsname\relax\def\urlprefix{URL }\fi
\providecommand{\bibinfo}[2]{#2}
\providecommand{\eprint}[2][]{\url{#2}}

\bibitem{kruglyak2010}
\bibinfo{author}{Kruglyak, V.~V.}, \bibinfo{author}{Demokritov, S.~O.} \&
  \bibinfo{author}{Grundler, D.}
\newblock \bibinfo{title}{Magnonics}.
\newblock \emph{\bibinfo{journal}{J. Phys. D: Appl. Phys.}}
  \textbf{\bibinfo{volume}{43}}, \bibinfo{pages}{264001}
  (\bibinfo{year}{2010}).

\bibitem{lenk2011}
\bibinfo{author}{Lenk, B.}, \bibinfo{author}{Ulrichs, H.},
  \bibinfo{author}{Garbs, F.} \& \bibinfo{author}{M{\"u}nzenberg, M.}
\newblock \bibinfo{title}{The building blocks of magnonics}.
\newblock \emph{\bibinfo{journal}{Phys. Rep.}} \textbf{\bibinfo{volume}{507}},
  \bibinfo{pages}{107--136} (\bibinfo{year}{2011}).

\bibitem{khitun2010}
\bibinfo{author}{Khitun, A.}, \bibinfo{author}{Bao, M.} \&
  \bibinfo{author}{Wang, K.~L.}
\newblock \bibinfo{title}{Magnonic logic circuits}.
\newblock \emph{\bibinfo{journal}{J. Phys D: Appl. Phys.}}
  \textbf{\bibinfo{volume}{43}}, \bibinfo{pages}{264005}
  (\bibinfo{year}{2010}).

\bibitem{kajiwara2010}
\bibinfo{author}{Kajiwara, Y.} \emph{et~al.}
\newblock \bibinfo{title}{Transmission of electrical signals by spin-wave
  interconversion in a magnetic insulator}.
\newblock \emph{\bibinfo{journal}{Nature}} \textbf{\bibinfo{volume}{464}},
  \bibinfo{pages}{262--266} (\bibinfo{year}{2010}).

\bibitem{chumak2015}
\bibinfo{author}{Chumak, A.~V.}, \bibinfo{author}{Vasyuchka, V.~I.},
  \bibinfo{author}{Serga, A.~A.} \& \bibinfo{author}{Hillebrands, B.}
\newblock \bibinfo{title}{Magnon spintronics}.
\newblock \emph{\bibinfo{journal}{Nature Phys.}} \textbf{\bibinfo{volume}{11}},
  \bibinfo{pages}{453--461} (\bibinfo{year}{2015}).

\bibitem{bloch1930}
\bibinfo{author}{Bloch, F.}
\newblock \bibinfo{title}{Zur theorie des ferromagnetismus}.
\newblock \emph{\bibinfo{journal}{Zeitschrift f{\"u}r Physik}}
  \textbf{\bibinfo{volume}{61}}, \bibinfo{pages}{206--219}
  (\bibinfo{year}{1930}).

\bibitem{han2019}
\bibinfo{author}{Han, J.}, \bibinfo{author}{Zhang, P.}, \bibinfo{author}{Hou,
  J.~T.}, \bibinfo{author}{Siddiqui, S.~A.} \& \bibinfo{author}{Liu, L.}
\newblock \bibinfo{title}{Mutual control of coherent spin waves and magnetic
  domain walls in a magnonic device}.
\newblock \emph{\bibinfo{journal}{Science}} \textbf{\bibinfo{volume}{366}},
  \bibinfo{pages}{1121--1125} (\bibinfo{year}{2019}).

\bibitem{hamadeh2014}
\bibinfo{author}{Hamadeh, A.} \emph{et~al.}
\newblock \bibinfo{title}{Full control of the spin-wave damping in a magnetic
  insulator using spin-orbit torque}.
\newblock \emph{\bibinfo{journal}{Phys. Rev. Lett.}}
  \textbf{\bibinfo{volume}{113}}, \bibinfo{pages}{197203}
  (\bibinfo{year}{2014}).

\bibitem{zhou2022}
\bibinfo{author}{Zhou, Y.} \emph{et~al.}
\newblock \bibinfo{title}{Piezoelectric strain-controlled magnon spin current
  transport in an antiferromagnet}.
\newblock \emph{\bibinfo{journal}{Nano Lett.}}  (\bibinfo{year}{2022}).

\bibitem{demokritov2004}
\bibinfo{author}{Demokritov, S.~O.} \emph{et~al.}
\newblock \bibinfo{title}{Tunneling of dipolar spin waves through a region of
  inhomogeneous magnetic field}.
\newblock \emph{\bibinfo{journal}{Phys. Rev. Lett.}}
  \textbf{\bibinfo{volume}{93}}, \bibinfo{pages}{047201}
  (\bibinfo{year}{2004}).

\bibitem{wang2014}
\bibinfo{author}{Wang, H.}, \bibinfo{author}{Du, C.}, \bibinfo{author}{Hammel,
  P.~C.} \& \bibinfo{author}{Yang, F.}
\newblock \bibinfo{title}{{Antiferromagnonic spin transport from
  Y$_3$Fe$_5$O$_{12}$ into NiO}}.
\newblock \emph{\bibinfo{journal}{Phys. Rev. Lett.}}
  \textbf{\bibinfo{volume}{113}}, \bibinfo{pages}{097202}
  (\bibinfo{year}{2014}).

\bibitem{chen2018}
\bibinfo{author}{Chen, X.~Z.} \emph{et~al.}
\newblock \bibinfo{title}{Antidamping-torque-induced switching in biaxial
  antiferromagnetic insulators}.
\newblock \emph{\bibinfo{journal}{Phys. Rev. Lett.}}
  \textbf{\bibinfo{volume}{120}}, \bibinfo{pages}{207204}
  (\bibinfo{year}{2018}).

\bibitem{wang2003}
\bibinfo{author}{Wang, J.} \emph{et~al.}
\newblock \bibinfo{title}{Epitaxial {B}i{F}e{O}$_3$ multiferroic thin film
  heterostructures}.
\newblock \emph{\bibinfo{journal}{Science}} \textbf{\bibinfo{volume}{299}},
  \bibinfo{pages}{1719--1722} (\bibinfo{year}{2003}).

\bibitem{wang2019}
\bibinfo{author}{Wang, Y.} \emph{et~al.}
\newblock \bibinfo{title}{Magnetization switching by magnon-mediated spin
  torque through an antiferromagnetic insulator}.
\newblock \emph{\bibinfo{journal}{Science}} \textbf{\bibinfo{volume}{366}},
  \bibinfo{pages}{1125--1128} (\bibinfo{year}{2019}).

\bibitem{lebrun2018}
\bibinfo{author}{Lebrun, R.} \emph{et~al.}
\newblock \bibinfo{title}{Tunable long-distance spin transport in a crystalline
  antiferromagnetic iron oxide}.
\newblock \emph{\bibinfo{journal}{Nature}} \textbf{\bibinfo{volume}{561}},
  \bibinfo{pages}{222--225} (\bibinfo{year}{2018}).

\bibitem{wang2019spin}
\bibinfo{author}{Wang, H.} \emph{et~al.}
\newblock \bibinfo{title}{Spin-orbit-torque switching mediated by an
  antiferromagnetic insulator}.
\newblock \emph{\bibinfo{journal}{Phys. Rev. Appl.}}
  \textbf{\bibinfo{volume}{11}}, \bibinfo{pages}{044070}
  (\bibinfo{year}{2019}).

\bibitem{chen2019}
\bibinfo{author}{Chen, X.} \emph{et~al.}
\newblock \bibinfo{title}{Electric field control of {N{\'e}el} spin--orbit
  torque in an antiferromagnet}.
\newblock \emph{\bibinfo{journal}{Nature Mater.}}
  \textbf{\bibinfo{volume}{18}}, \bibinfo{pages}{931--935}
  (\bibinfo{year}{2019}).

\bibitem{han2020}
\bibinfo{author}{Han, J.} \emph{et~al.}
\newblock \bibinfo{title}{Birefringence-like spin transport via linearly
  polarized antiferromagnetic magnons}.
\newblock \emph{\bibinfo{journal}{Nature Nanotechnol.}}
  \textbf{\bibinfo{volume}{15}}, \bibinfo{pages}{563--568}
  (\bibinfo{year}{2020}).

\bibitem{dkabrowski2020}
\bibinfo{author}{Dkabrowski, M.} \emph{et~al.}
\newblock \bibinfo{title}{Coherent transfer of spin angular momentum by
  evanescent spin waves within antiferromagnetic nio}.
\newblock \emph{\bibinfo{journal}{Phys. Rev. Lett.}}
  \textbf{\bibinfo{volume}{124}}, \bibinfo{pages}{217201}
  (\bibinfo{year}{2020}).

\bibitem{chen2021}
\bibinfo{author}{Chen, X.} \emph{et~al.}
\newblock \bibinfo{title}{Observation of the antiferromagnetic spin {Hall}
  effect}.
\newblock \emph{\bibinfo{journal}{Nature Mater.}}
  \textbf{\bibinfo{volume}{20}}, \bibinfo{pages}{800--804}
  (\bibinfo{year}{2021}).

\bibitem{zhang2022control}
\bibinfo{author}{Zhang, P.} \emph{et~al.}
\newblock \bibinfo{title}{Control of n{\'e}el vector with spin-orbit torques in
  an antiferromagnetic insulator with tilted easy plane}.
\newblock \emph{\bibinfo{journal}{Phys. Rev. Lett.}}
  \textbf{\bibinfo{volume}{129}}, \bibinfo{pages}{017203}
  (\bibinfo{year}{2022}).

\bibitem{moon2013}
\bibinfo{author}{Moon, J.-H.} \emph{et~al.}
\newblock \bibinfo{title}{Spin-wave propagation in the presence of interfacial
  {Dzyaloshinskii-Moriya} interaction}.
\newblock \emph{\bibinfo{journal}{Phys. Rev. B}} \textbf{\bibinfo{volume}{88}},
  \bibinfo{pages}{184404} (\bibinfo{year}{2013}).

\bibitem{iguchi2015}
\bibinfo{author}{Iguchi, Y.}, \bibinfo{author}{Uemura, S.},
  \bibinfo{author}{Ueno, K.} \& \bibinfo{author}{Onose, Y.}
\newblock \bibinfo{title}{Nonreciprocal magnon propagation in a
  noncentrosymmetric ferromagnet {LiFe$_5$O$_8$}}.
\newblock \emph{\bibinfo{journal}{Phys. Rev. B}} \textbf{\bibinfo{volume}{92}},
  \bibinfo{pages}{184419} (\bibinfo{year}{2015}).

\bibitem{gitgeatpong2017}
\bibinfo{author}{Gitgeatpong, G.} \emph{et~al.}
\newblock \bibinfo{title}{Nonreciprocal magnons and symmetry-breaking in the
  noncentrosymmetric antiferromagnet}.
\newblock \emph{\bibinfo{journal}{Phys. Rev. Lett.}}
  \textbf{\bibinfo{volume}{119}}, \bibinfo{pages}{047201}
  (\bibinfo{year}{2017}).

\bibitem{kimura2003}
\bibinfo{author}{Kimura, T.} \emph{et~al.}
\newblock \bibinfo{title}{Magnetic control of ferroelectric polarization}.
\newblock \emph{\bibinfo{journal}{Nature}} \textbf{\bibinfo{volume}{426}},
  \bibinfo{pages}{55--58} (\bibinfo{year}{2003}).

\bibitem{Chu2008}
\bibinfo{author}{Chu, Y.-H.} \emph{et~al.}
\newblock \bibinfo{title}{Electric-field control of local ferromagnetism using
  a magnetoelectric multiferroic}.
\newblock \emph{\bibinfo{journal}{Nature Mater.}} \textbf{\bibinfo{volume}{7}},
  \bibinfo{pages}{478} (\bibinfo{year}{2008}).

\bibitem{Zhao2006}
\bibinfo{author}{Zhao, T.} \emph{et~al.}
\newblock \bibinfo{title}{Electrical control of antiferromagnetic domains in
  multiferroic {BiFeO$_3$} films at room temperature}.
\newblock \emph{\bibinfo{journal}{Nature Mater.}} \textbf{\bibinfo{volume}{5}},
  \bibinfo{pages}{823} (\bibinfo{year}{2006}).

\bibitem{rovillain2010}
\bibinfo{author}{Rovillain, P.} \emph{et~al.}
\newblock \bibinfo{title}{Electric-field control of spin waves at room
  temperature in multiferroic {Bi}{Fe}{O}$_3$}.
\newblock \emph{\bibinfo{journal}{Nature Mater.}} \textbf{\bibinfo{volume}{9}},
  \bibinfo{pages}{975--979} (\bibinfo{year}{2010}).

\bibitem{gross2017real}
\bibinfo{author}{Gross, I.} \emph{et~al.}
\newblock \bibinfo{title}{Real-space imaging of non-collinear antiferromagnetic
  order with a single-spin magnetometer}.
\newblock \emph{\bibinfo{journal}{Nature}} \textbf{\bibinfo{volume}{549}},
  \bibinfo{pages}{252--256} (\bibinfo{year}{2017}).

\bibitem{sando2013}
\bibinfo{author}{Sando, D.} \emph{et~al.}
\newblock \bibinfo{title}{Crafting the magnonic and spintronic response of
  {Bi}{Fe}{O}$_3$ films by epitaxial strain}.
\newblock \emph{\bibinfo{journal}{Nature Mater.}}
  \textbf{\bibinfo{volume}{12}}, \bibinfo{pages}{641--646}
  (\bibinfo{year}{2013}).

\bibitem{Fischer1980}
\bibinfo{author}{Fischer, P.}, \bibinfo{author}{Polomska, M.},
  \bibinfo{author}{Sosnowska, I.} \& \bibinfo{author}{Szymanski, M.}
\newblock \bibinfo{title}{Temperature dependence of the crystal and magnetic
  structure of {BiFeO$_3$}}.
\newblock \emph{\bibinfo{journal}{J. Phys. C: Solid St. Phys.}}
  \textbf{\bibinfo{volume}{19}}, \bibinfo{pages}{1931} (\bibinfo{year}{1980}).

\bibitem{haykal2020}
\bibinfo{author}{Haykal, A.} \emph{et~al.}
\newblock \bibinfo{title}{Antiferromagnetic textures in {B}i{F}e{O}$_3$
  controlled by strain and electric field}.
\newblock \emph{\bibinfo{journal}{Nature Communs.}}
  \textbf{\bibinfo{volume}{11}}, \bibinfo{pages}{1--7} (\bibinfo{year}{2020}).

\bibitem{Ederer2005}
\bibinfo{author}{Ederer, C.} \& \bibinfo{author}{Spaldin, N.~A.}
\newblock \bibinfo{title}{Weak ferromagnetism and magnetoelectric coupling in
  bismuth ferrite}.
\newblock \emph{\bibinfo{journal}{Phys. Rev. B}} \textbf{\bibinfo{volume}{71}},
  \bibinfo{pages}{060401(R)} (\bibinfo{year}{2005}).

\bibitem{Neaton2005}
\bibinfo{author}{Neaton, J.~B.}, \bibinfo{author}{Ederer, C.},
  \bibinfo{author}{Waghmare, U.~V.}, \bibinfo{author}{Spaldin, N.~A.} \&
  \bibinfo{author}{Rabe, K.~M.}
\newblock \bibinfo{title}{First-principles study of spontaneous polarization in
  multiferroic {BiFeO$_3$}}.
\newblock \emph{\bibinfo{journal}{Phys. Rev. B}} \textbf{\bibinfo{volume}{71}},
  \bibinfo{pages}{014113} (\bibinfo{year}{2005}).

\bibitem{Heron2014}
\bibinfo{author}{Heron, J.} \emph{et~al.}
\newblock \bibinfo{title}{Deterministic switching of ferromagnetism at room
  temperature using an electric field}.
\newblock \emph{\bibinfo{journal}{Nature}} \textbf{\bibinfo{volume}{516}},
  \bibinfo{pages}{370--373} (\bibinfo{year}{2014}).

\bibitem{Rahmedov2012}
\bibinfo{author}{Rahmedov, D.}, \bibinfo{author}{Wang, D.},
  \bibinfo{author}{Íñiguez, J.} \& \bibinfo{author}{Bellaiche, L.}
\newblock \bibinfo{title}{Magnetic cycloid of {B}i{F}e{O}$_3$ from atomistic
  simulations}.
\newblock \emph{\bibinfo{journal}{Phys. Rev. Lett.}}
  \textbf{\bibinfo{volume}{109}}, \bibinfo{pages}{037207}
  (\bibinfo{year}{2012}).

\bibitem{competingDMI}
\bibinfo{author}{Ramazanoglu, M.} \emph{et~al.}
\newblock \bibinfo{title}{Local weak ferromagnetism in single-crystalline
  ferroelectric {B}i{F}e{O}$_3$}.
\newblock \emph{\bibinfo{journal}{Phys. Rev. Lett.}}
  \textbf{\bibinfo{volume}{107}}, \bibinfo{pages}{207206}
  (\bibinfo{year}{2011}).

\bibitem{yang2010}
\bibinfo{author}{Yang, S.} \emph{et~al.}
\newblock \bibinfo{title}{Above-bandgap voltages from ferroelectric
  photovoltaic devices}.
\newblock \emph{\bibinfo{journal}{Nature Nanotechnol.}}
  \textbf{\bibinfo{volume}{5}}, \bibinfo{pages}{143--147}
  (\bibinfo{year}{2010}).

\bibitem{wang2020}
\bibinfo{author}{Wang, H.} \emph{et~al.}
\newblock \bibinfo{title}{Chiral spin-wave velocities induced by all-garnet
  interfacial {Dzyaloshinskii-Moriya} interaction in ultrathin yttrium iron
  garnet films}.
\newblock \emph{\bibinfo{journal}{Phys. Rev. Lett.}}
  \textbf{\bibinfo{volume}{124}}, \bibinfo{pages}{027203}
  (\bibinfo{year}{2020}).

\bibitem{huang2021novel}
\bibinfo{author}{Huang, X.} \emph{et~al.}
\newblock \bibinfo{title}{Novel spin--orbit torque generation at room
  temperature in an all-oxide epitaxial {L}a$_{0. 7}${S}r$_{0.
  3}${M}n{O}$_3$/{S}r{I}r{O}$_3$ system}.
\newblock \emph{\bibinfo{journal}{Adv. Mater.}} \textbf{\bibinfo{volume}{33}},
  \bibinfo{pages}{2008269} (\bibinfo{year}{2021}).

\bibitem{huang2020}
\bibinfo{author}{Huang, Y.-L.} \emph{et~al.}
\newblock \bibinfo{title}{Manipulating magnetoelectric energy landscape in
  multiferroics}.
\newblock \emph{\bibinfo{journal}{Nature Commun.}}
  \textbf{\bibinfo{volume}{11}}, \bibinfo{pages}{2836} (\bibinfo{year}{2020}).

\bibitem{manipatruni2019}
\bibinfo{author}{Manipatruni, S.} \emph{et~al.}
\newblock \bibinfo{title}{Scalable energy-efficient magnetoelectric spin--orbit
  logic}.
\newblock \emph{\bibinfo{journal}{Nature}} \textbf{\bibinfo{volume}{565}},
  \bibinfo{pages}{35--42} (\bibinfo{year}{2019}).

\bibitem{noel2020}
\bibinfo{author}{No{\"e}l, P.} \emph{et~al.}
\newblock \bibinfo{title}{Non-volatile electric control of spin--charge
  conversion in a {Sr}{Ti}{O}$_3$ {R}ashba system}.
\newblock \emph{\bibinfo{journal}{Nature}} \textbf{\bibinfo{volume}{580}},
  \bibinfo{pages}{483--486} (\bibinfo{year}{2020}).

\bibitem{varotto2021}
\bibinfo{author}{Varotto, S.} \emph{et~al.}
\newblock \bibinfo{title}{Room-temperature ferroelectric switching of
  spin-to-charge conversion in germanium telluride}.
\newblock \emph{\bibinfo{journal}{Nature Electron.}}
  \textbf{\bibinfo{volume}{4}}, \bibinfo{pages}{740--747}
  (\bibinfo{year}{2021}).

\bibitem{lesne2016}
\bibinfo{author}{Lesne, E.} \emph{et~al.}
\newblock \bibinfo{title}{Highly efficient and tunable spin-to-charge
  conversion through {R}ashba coupling at oxide interfaces}.
\newblock \emph{\bibinfo{journal}{Nature Mater.}}
  \textbf{\bibinfo{volume}{15}}, \bibinfo{pages}{1261--1266}
  (\bibinfo{year}{2016}).

\bibitem{vaz2019}
\bibinfo{author}{Vaz, D.~C.} \emph{et~al.}
\newblock \bibinfo{title}{Mapping spin--charge conversion to the band structure
  in a topological oxide two-dimensional electron gas}.
\newblock \emph{\bibinfo{journal}{Nature Mater.}}
  \textbf{\bibinfo{volume}{18}}, \bibinfo{pages}{1187--1193}
  (\bibinfo{year}{2019}).

\bibitem{liu2012}
\bibinfo{author}{Liu, L.} \emph{et~al.}
\newblock \bibinfo{title}{Spin-torque switching with the giant spin {Hall}
  effect of tantalum}.
\newblock \emph{\bibinfo{journal}{Science}} \textbf{\bibinfo{volume}{336}},
  \bibinfo{pages}{555--558} (\bibinfo{year}{2012}).

\bibitem{merbouche2021}
\bibinfo{author}{Merbouche, H.} \emph{et~al.}
\newblock \bibinfo{title}{Voltage-controlled reconfigurable magnonic crystal at
  the sub-micrometer scale}.
\newblock \emph{\bibinfo{journal}{ACS Nano}} \textbf{\bibinfo{volume}{15}},
  \bibinfo{pages}{9775--9781} (\bibinfo{year}{2021}).

\bibitem{yi2019}
\bibinfo{author}{Yi, D.} \emph{et~al.}
\newblock \bibinfo{title}{Tailoring magnetoelectric coupling in
  {B}i{F}e{O}$_3$/{L}a$_{0. 7}${S}r$_{0. 3}${M}n{O}$_3$ heterostructure through
  the interface engineering}.
\newblock \emph{\bibinfo{journal}{Adv. Mater.}} \textbf{\bibinfo{volume}{31}},
  \bibinfo{pages}{1806335} (\bibinfo{year}{2019}).

\bibitem{yu2012}
\bibinfo{author}{Yu, P.} \emph{et~al.}
\newblock \bibinfo{title}{Interface control of bulk ferroelectric
  polarization}.
\newblock \emph{\bibinfo{journal}{PNAS}} \textbf{\bibinfo{volume}{109}},
  \bibinfo{pages}{9710--9715} (\bibinfo{year}{2012}).

\bibitem{parsonnet2022}
\bibinfo{author}{Parsonnet, E.} \emph{et~al.}
\newblock \bibinfo{title}{Non-volatile electric field control of thermal
  magnons in the absence of an applied magnetic field}.
\newblock \emph{\bibinfo{journal}{Phys. Rev. Lett.}}
  \textbf{\bibinfo{volume}{129}}, \bibinfo{pages}{087601}
  (\bibinfo{year}{2022}).

\bibitem{wang2019large}
\bibinfo{author}{Wang, H.} \emph{et~al.}
\newblock \bibinfo{title}{Large spin-orbit torque observed in epitaxial
  {S}r{I}r{O}$_3$ thin films}.
\newblock \emph{\bibinfo{journal}{Appl. Phys. Lett.}}
  \textbf{\bibinfo{volume}{114}}, \bibinfo{pages}{232406}
  (\bibinfo{year}{2019}).

\bibitem{cornelissen2015}
\bibinfo{author}{Cornelissen, L.~J.}, \bibinfo{author}{Liu, J.},
  \bibinfo{author}{Duine, R.~A.}, \bibinfo{author}{Youssef, J.~B.} \&
  \bibinfo{author}{Van~Wees, B.~J.}
\newblock \bibinfo{title}{Long-distance transport of magnon spin information in
  a magnetic insulator at room temperature}.
\newblock \emph{\bibinfo{journal}{Nature Phys.}} \textbf{\bibinfo{volume}{11}},
  \bibinfo{pages}{1022--1026} (\bibinfo{year}{2015}).

\bibitem{shikoh2013}
\bibinfo{author}{Shikoh, E.} \emph{et~al.}
\newblock \bibinfo{title}{Spin-pump-induced spin transport in p-type {Si} at
  room temperature}.
\newblock \emph{\bibinfo{journal}{Phys. Rev. Lett.}}
  \textbf{\bibinfo{volume}{110}}, \bibinfo{pages}{127201}
  (\bibinfo{year}{2013}).

\bibitem{nan2019}
\bibinfo{author}{Nan, T.} \emph{et~al.}
\newblock \bibinfo{title}{Anisotropic spin-orbit torque generation in epitaxial
  {SrIrO$_3$} by symmetry design}.
\newblock \emph{\bibinfo{journal}{PNAS}} \textbf{\bibinfo{volume}{116}},
  \bibinfo{pages}{16186--16191} (\bibinfo{year}{2019}).

\bibitem{Cheng2014}
\bibinfo{author}{Cheng, R.}, \bibinfo{author}{Xiao, J.}, \bibinfo{author}{Niu,
  Q.} \& \bibinfo{author}{Brataas, A.}
\newblock \bibinfo{title}{Spin pumping and spin-transfer torques in
  antiferromagnets}.
\newblock \emph{\bibinfo{journal}{Phys. Rev. Lett.}}
  \textbf{\bibinfo{volume}{113}} (\bibinfo{year}{2014}).

\bibitem{Mangeri2017}
\bibinfo{author}{Mangeri, J.} \emph{et~al.}
\newblock \bibinfo{title}{Topological phase transformations in ferroelectric
  nanoparticles}.
\newblock \emph{\bibinfo{journal}{Nanoscale}} \textbf{\bibinfo{volume}{9}},
  \bibinfo{pages}{1616--1624} (\bibinfo{year}{2017}).

\bibitem{Permann2020}
\bibinfo{author}{Permann, C.~J.} \emph{et~al.}
\newblock \bibinfo{title}{Moose: Enabling massively parallel multiphysics
  simulation}.
\newblock \emph{\bibinfo{journal}{SoftwareX}} \textbf{\bibinfo{volume}{11}},
  \bibinfo{pages}{100430} (\bibinfo{year}{2020}).

\bibitem{liu2011}
\bibinfo{author}{Liu, L.}, \bibinfo{author}{Moriyama, T.},
  \bibinfo{author}{Ralph, D.~C.} \& \bibinfo{author}{Buhrman, R.~A.}
\newblock \bibinfo{title}{Spin-torque ferromagnetic resonance induced by the
  spin {Hall} effect}.
\newblock \emph{\bibinfo{journal}{Phys. Rev. Lett.}}
  \textbf{\bibinfo{volume}{106}}, \bibinfo{pages}{036601}
  (\bibinfo{year}{2011}).

\bibitem{pai2015}
\bibinfo{author}{Pai, C.-F.}, \bibinfo{author}{Ou, Y.},
  \bibinfo{author}{Vilela-Leao, L.~H.}, \bibinfo{author}{Ralph, D.~C.} \&
  \bibinfo{author}{Buhrman, R.~A.}
\newblock \bibinfo{title}{Dependence of the efficiency of spin {Hall} torque on
  the transparency of {Pt}/ferromagnetic layer interfaces}.
\newblock \emph{\bibinfo{journal}{Phys. Rev. B}} \textbf{\bibinfo{volume}{92}},
  \bibinfo{pages}{064426} (\bibinfo{year}{2015}).

\bibitem{wang2015}
\bibinfo{author}{Wang, Y.} \emph{et~al.}
\newblock \bibinfo{title}{Topological surface states originated spin-orbit
  torques in {Bi$_2$ Se$_3$}}.
\newblock \emph{\bibinfo{journal}{Phys. Rev. Lett.}}
  \textbf{\bibinfo{volume}{114}}, \bibinfo{pages}{257202}
  (\bibinfo{year}{2015}).

\bibitem{rijnders2004}
\bibinfo{author}{Rijnders, G.}, \bibinfo{author}{Blank, D.~H.},
  \bibinfo{author}{Choi, J.} \& \bibinfo{author}{Eom, C.-B.}
\newblock \bibinfo{title}{Enhanced surface diffusion through termination
  conversion during epitaxial {Sr}{Ru}{O}$_3$ growth}.
\newblock \emph{\bibinfo{journal}{Appl. Phys. Lett.}}
  \textbf{\bibinfo{volume}{84}}, \bibinfo{pages}{505--507}
  (\bibinfo{year}{2004}).

\end{thebibliography}
    
    \noindent\textbf{Methods}
\\
    \noindent\textbf{Sample and device fabrication}\\
    The tri-layers used in this study were grown by RHEED assisted pulsed laser deposition using a KrF laser with a wavelength of 248 nm. A 20 nm La$_{0.7}$Sr$_{0.3}$MnO$_3$ {\color{blue}{(5 nm SrRuO$_3$)}}layer was first grown at a substrate temperature 700$^\circ$C and oxygen partial pressure 150 mTorr {\color{blue}{(100 mTorr)}}, then a BiFeO$_3$ layer followed with various thicknesses at a substrate temperature 700$^\circ$C and oxygen partial pressure 100 mTorr, and finally a 10 nm SrIrO$_3$ layer finished the growth at a substrate temperature 700$^\circ$C and oxygen partial pressure 50 mTorr. The films were grown at a repetition of 2 {\color{blue}{(5 Hz)}} Hz, 10 Hz and 5 Hz with a laser fluence of 1.5 J/cm$^2$, 1.5 J/cm$^2$ and 1.2 J/cm$^2$, respectively. After the growth, the sample was cooled to room temperature in an oxygen environment of 750 Torr at a cooling rate of 10$^\circ$C/min.
    To make ST-FMR devices, the samples were first spin coated with photoresist (MiR 701), then devices with a $50\times 25$ $\mu$m$^2$ rectangular dimensions ($15 \times 10$ $\mu$m$^2$ for the electric-field-controlled experiment)  were made by photolithography (Heidelberg MLA 150) followed by ion milling (Pi scientific) that stopped at the La$_{0.7}$Sr$_{0.3}$MnO$_3$ layer. The remaining La$_{0.7}$Sr$_{0.3}$MnO$_3$ layer was used as the bottom electrode in the switching experiments; for the ST-FMR measurements this was patterned using a H$_3$PO$_4$ wet etch (H$_3$PO$_4$ : H$_2$O = 1 : 3). Finally, electrical contacts were made from 100 nm of Pt with a ground-signal-ground (GSG) geometry , so that when a GSG high-frequency probe made into contact with the samples the current traveling through the Pt contacts did not produce a net Oersted field at the sample.\\
\\
    \noindent\textbf{Antiferromagnetic spin wave calculations}
\\
To perform the spin wave calculations, we first prepare a polar ground state corresponding to $\mathbf{P}||[111]$ and $\mathbf{P}||[\bar{1}\bar{1}\bar{1}]$ (along with an order parameter for the octahedral antiphase tilts). We then evolve the LLG-LLB equation to find the magnetic ground state corresponding to $\mathbf{L}||[\bar{1}01]$ with $\mathbf{m}||[1\bar{2}1]$ and $\mathbf{m}||[\bar{1}2\bar{1}]$. 
To study the spin waves, the homogeneous magnetic ground state is perturbed with an applied field $\mathbf{H}||[\bar{1}01]$ with Gilbert damping $\alpha = 0$, and the spin waves are set to propagate along $\pm\mathbf{k}||[001]$.
The detectable spin current in the AFM insulating layer\cite{Cheng2014} are proportional to $\mathbf{j} = \mathbf{j}^\mathbf{l} + \mathbf{j}^\mathbf{m} + \mathbf{j}^{\mathbf{ml}}+\mathbf{j}^{\mathbf{lm}}$ where $\mathbf{j}^\mathbf{l} \propto \mathbf{l}\times\dot{\mathbf{l}}$, $\mathbf{j}^\mathbf{m} \propto \mathbf{m}\times\dot{\mathbf{m}}$,
$\mathbf{j}^\mathbf{ml} \propto\mathbf{m}\times\dot{\mathbf{l}}$, and 
$\mathbf{j}^\mathbf{lm} \propto \mathbf{l}\times\dot{\mathbf{m}}$.
More details of our simulations are provided in the SM.
Calculations are performed using the Ferret module \cite{Mangeri2017} which is part of the open-source MOOSE framework \cite{Permann2020}.\\

    \noindent\textbf{ST-FMR analysis}\\
    The ST-FMR signal, $V_{mix}$, is produced across the La$_{0.7}$Sr$_{0.3}$MnO$_3$/BiFeO$_3$/SrIrO$_3$ ST-FMR device as rectification of anisotropic magnetoresistance that oscillates at the same frequency as the input microwave current $I_{RF}$\upcite{liu2011}. A microwave current at frequencies 4-6 GHz and power 12 dBm is applied to the device, with an in-plane external magnetic field oriented 45$^\circ$ with respect to the current direction. Finally, \(V_{mix}=-\dfrac{1}{4}\dfrac{dR}{d\theta}\dfrac{\gamma I_{RF} cos\theta}{2\pi \Delta (df/dH)_{H_{ext}=H_0}}(\tau_{DL} F_S(H_{ext}) +\tau_{FL} F_A(H_{ext}))\)\upcite{liu2011}, where \(dR/d\theta\) is the angle dependent magnetoresistance at $\theta$,  $\gamma$ is the gyromagnetic ratio, $\Delta$ is the linewidth of the ST-FMR signal, \((df⁄dH)_{H_{ext}=H_0}\) is the field gradient of the resonance frequency, $\tau_{DL}$ is the damping-like torque, $\tau_{FL}$ is the field-like torque, $F_S (H_{ext} )$ is the symmetric Lorentzian function and $F_A (H_{ext} )$ is the anti-symmetric Lorentzian function respectively, is detected by Keithley 2182A nanovoltmeter, from the fits of which the symmetric component $V_S$ and the anti-symmetric component $V_A$, which are associated with damping-like torque ($\tau_{DL}$) and field-like torque ($\tau_{FL}$) respectively, can be extracted. As a result, spin orbit torque efficiency-$\xi_{SOT}$ is evaluated using \(\xi_{DL}=\dfrac{2e\tau_{DL} m_s t_{FM}}{J_c\hbar}\), where \(\tau_{DL}=\dfrac{4V_s \Delta}{I_{RF} cos\theta\dfrac{dR}{d\theta}}\dfrac{1+\dfrac{\mu_0 M_{eff}}{2B_{res}}}{(\dfrac{1+\mu_0 M_{eff}}{B_{res}})^{1/2}}\)\upcite{pai2015, wang2015}. The SOT efficiency for the samples in this work was measured from 4 GHz to 6 GHz and the error bar in Fig.~\ref{fig:2} is a result of SOT efficiencies across different frequencies, while the error bar in extended Fig.~\ref{fig:Eig2} is caused by the SOT efficiency distribution across different devices.\\
     \\  
    \\
    \noindent\textbf{\color{blue}{Parameters for non-local spin transport measurements}}\\
    {\color{blue}{The length of the wire is 100 $\mu m$, the applied voltage is ~ 20V, the applied current is 10 $\mu A$, the alternating current frequency is 7 Hz.}}
    \\
    \\
    \noindent\textbf{Data availability}\\
    The data that support the findings of this study are available from the corresponding authors on reasonable request.\\
\\ 
    \noindent\textbf{Acknowledgements}\\
    We are grateful for the fruitful discussions with Prof. Albert Fert, Dr. Eric Parsonnet, Dr. Yenlin Huang. X.H., D.C.R. and R.R. acknowledge the support from the SRC-JUMP ASCENT center. R.J. acknowledges the support from the U.S. Department of Energy, under contract No. DE-SC0017671. X.C., P.S., D.V., S.S., L.W.M., Z.Y., and R.R. acknowledge partial support from the U.S. Department of Energy, Office of Science, Office of Basic Energy Sciences, Materials Sciences and Engineering Division under Contract No. DE-AC02-05-CH11231 (Codesign of Ultra-Low-Voltage Beyond CMOS Microelectronics for the development of materials for low-power microelectronics). {\color{blue}{Y.L. and R.C. acknowledge the support from the Air Force Office of Scientific Research under Grant No. FA9550-19-1-0307.}}  T.W. and Z.Q.acknowledge the support from US Department of Energy, Office of Science, Office of Basic Energy Sciences, Materials Sciences and Engineering Division under Contract No. DE-AC02-05CH11231 (van der Waals heterostructures program, KCWF16). This research used resources of the Advanced Light Source, which is a DOE Office of Science User Facility under contract no. DE-AC02-05CH11231. H.P. acknowledges support from Army Research Office and Army Research Laboratory via the Collaborative for Hierarchical Agile and Responsive Materials (CHARM) under cooperative agreement W911NF-19-2-0119. J.M. has received funding from the European Union’s Horizon 2020 research and innovation programme under the Marie Skłodowska-Curie grant agreement SCALES - 897614. D.R.R. acknowledges funding from the Ministerio dell'Università e della Ricerca, Decreto Ministeriale n. 1062 del 10/08/2021 (PON Ricerca e Innovazione). O.H. acknowledges support from the US Department of Energy, Office of Science, Basic Energy Sciences, Division of Materials Sciences and Engineering.\\
\\    
    \noindent\textbf{Author contributions}\\
    X.C. and R.R. supervised this study. X.H. carried out the synthesis and characterization of heterostructures. X.H. and X.C. fabricated the devices. X.H. and R.J. carried out the ST-FMR measurements. X.H. and X.C. carried out the current induced magnetization switching measurements. C.X. {\color{blue}{H.T.}} performed the non-local transport measurements. C.K. performed dynamic XMCD measurements at beamline 4.0.2 of the Advanced Light Source. {\color{blue}{Y.L. and R.C performed the theoretical calculations.}} J.M., D.R.R. O.H., and J.I. developed and performed the coupled polar-micromagnetic simulations. X.C., D.V., and Z.Y. devised the electronic model and performed the calculations. S.S. performed the microstructure and electronic structure characterizations. H.Z., T.W., L.C., C.H., I.H., H.P., J.Y., P.M., P.S., Z.Q., S.S., M.R., D.S., L.W.M., D.C.R., and A.F. gave suggestions on the experiments. All authors discussed the results and prepared the manuscript.\\
 
    \noindent\textbf{Competing interests}\\
    The authors declare no competing interests.\\

    \noindent\textbf{Additional information}\\ Correspondence and requests for materials should be addressed to X.C. and R.R. 
    Reprints and permissions information is available at http://www.nature.com/reprints.

    \setcounter{figure}{0}
    \captionsetup[figure]{labelfont={bf}, name={Extended Data Fig.}, labelsep=period}

  \begin{figure}[t!]
    	\centering
    	\includegraphics[width=0.95\textwidth]{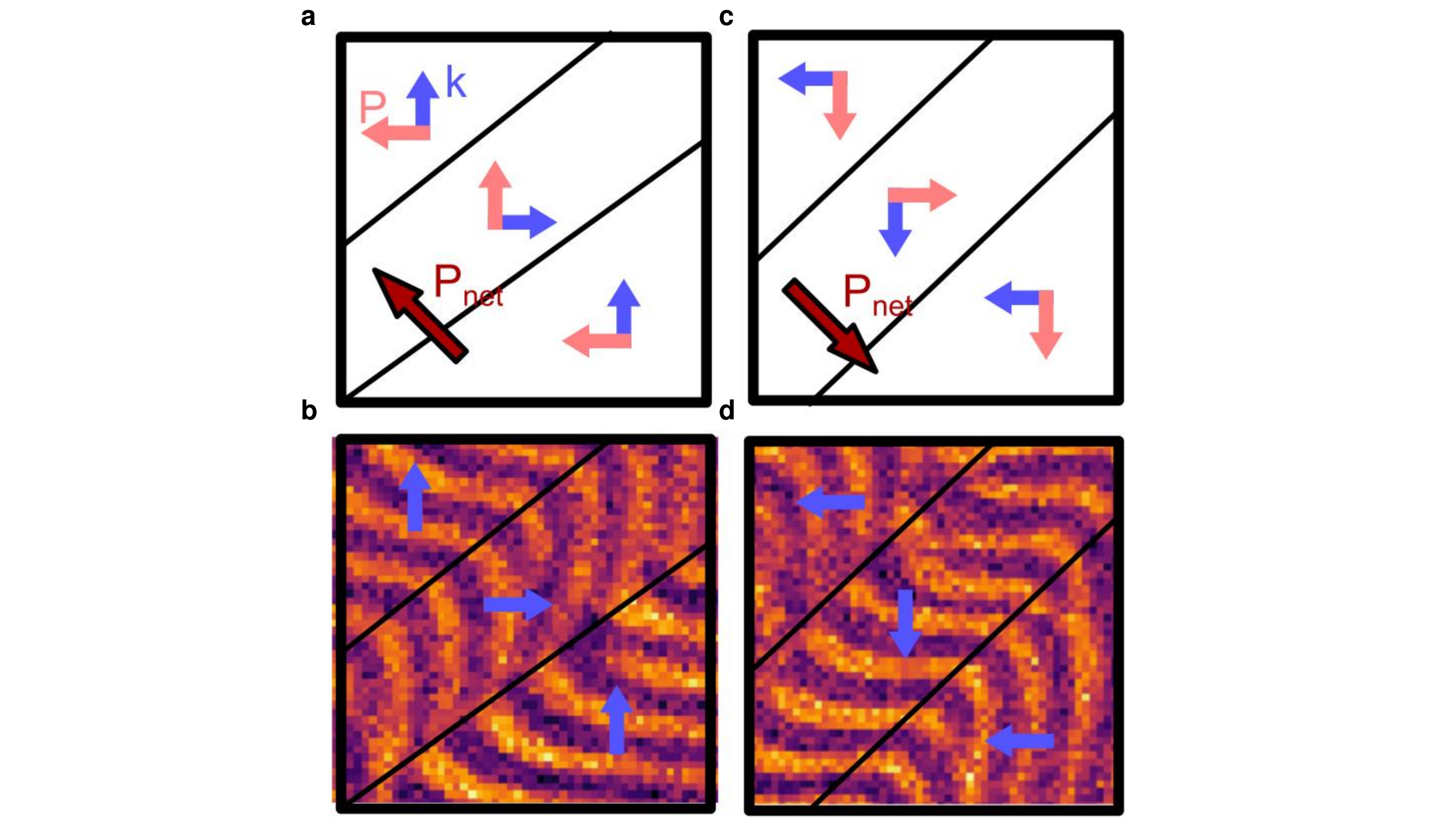}\caption{\color{blue}{\textbf{NV magnetometry image for BiFeO$_3$. a, c, }Schematic illustrations of ferroelectric domains and the corresponding spin cycloid propagation directions before and after polarization switching. The pink arrows represent the ferroelectric polarization orientations.\textbf{b, d, }Magnetic stray field distribution recorded with scanning NV magnetometer for 100 nm BiFeO$_3$ before and after polarization switching. The blue arrows are the spin cycloid propagation wave vector}}
    	\label{fig:Eig2}
    \end{figure}
     \begin{figure}[t!]
    	\centering
    	\includegraphics[width=1.0\textwidth]{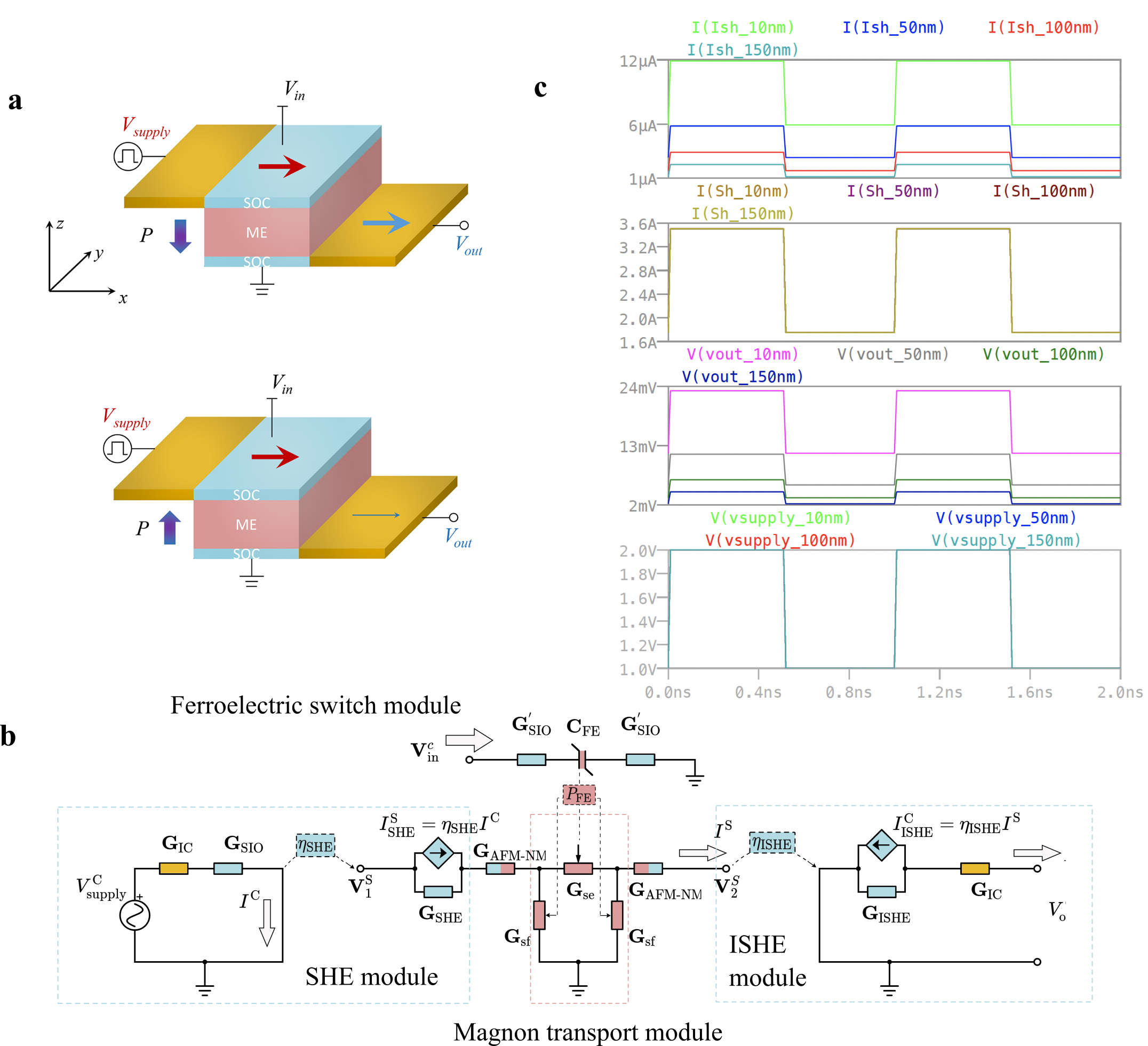}
    	\caption{\textbf{Schematics for magnon-mediated MESO logic device. 
    	a,} Same schematics duplicated from Fig.~\ref{fig:1}d. 
    	\textbf{b,} Modularized circuit schematics.  
    	The top module is the ferroelectric switch module, where the input voltage $V_\mathrm{in}^c$ switches the direction of the  ferroelectric polarization $P_\mathrm{FE}$ in the magnetoelectric layer.  
    	Such polarization change also affects the spin conductance  $G_\mathrm{se}$ and $G_\mathrm{sf}$ of the ME layer. 
    	The leftmost and rightmost modules represent the spin Hall effect (SHE) process in the top SOC layer (injector) and the inverse spin Hall effect (ISHE) process in the bottom SOC layer (detector), respectively. 
    	The current-controlled spin current source $I^S_\mathrm{SHE}$ in the SHE module depends on the output charge current of the top SOC layer. 
    	Similarly, the current-controlled charge current source $I^C_\mathrm{ISHE}$ in the ISHE module depends on the output spin current of the bottom SOC layer. 
    	The parameters $\eta_\textrm{SHE}$ and $\eta_\textrm{ISHE}$ represent the charge-to-spin and spin-to-charge current conversion rates, respectively.
    	The module in the middle describes the magnon transport process in the ME layer, which is connected to the SHE and ISHE modules through the spin conductance at the interfaces between ME and SOC layers, denoted as $G_\mathrm{AFM-NM}$. 
    	\textbf{c,} Lossy Buffer logic simulation result for the equivalent circuit in Fig.~\ref{fig:Efig5}.b where the output follows the input pulse signal characteristics.}
    	\label{fig:Efig5}
    \end{figure}
    
    \begin{figure}[t!]
    	\centering
    	\includegraphics[width=1\textwidth]{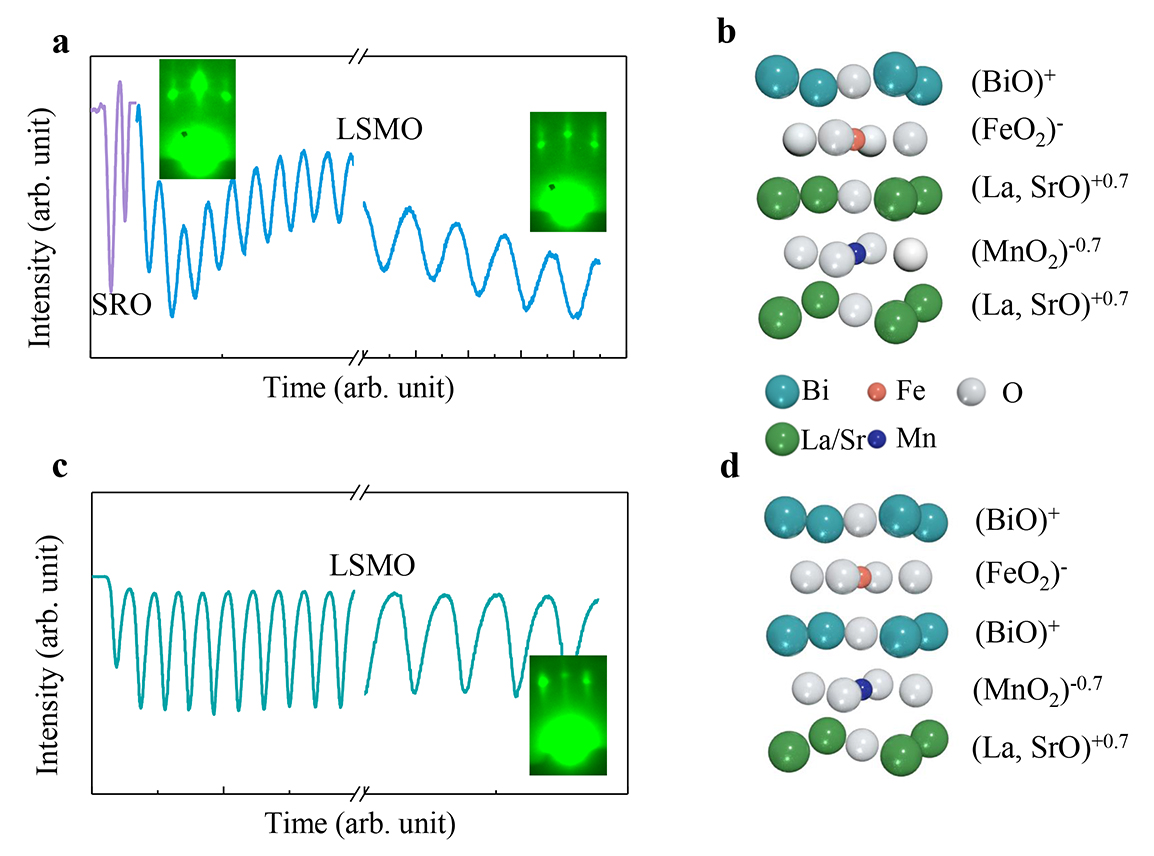}
    	\caption{\textbf{Atomic scale control of interface termination by pulsed laser deposition. a,} RHEED intensity oscillation of the specularly reflected beam during the growth of SrRuO$_3$ and La$_{0.7}$Sr$_{0.3}$MnO$_3$ on TiO$_2$-terminated SrTiO$_3$ $[0 0 1]$ surface. A growth of 2.5 unit-cell of SrRuO$_3$ on TiO$_2$ terminated SrTiO$_3$ substrate is required in order to achieve the SrO terminated substrate, due to the fact that RuO is highly volatile at high temperatures and SrRuo$_3$ is therefore self-terminated with SrO layer\upcite{rijnders2004}.\textbf{b,} A schematic of the resulting La$_{0.7}$Sr$_{0.3}$O-FeO$_2$ interface. \textbf{c,} RHEED intensity oscillation of the specular beam during the growth of La$_{0.7}$Sr$_{0.3}$MnO$_3$ by direct on TiO$_2$-terminated SrTiO$_3$ $[0 0 1]$ surface. \textbf{d,} A schematic of the resulting MnO$_2$-BiO interface.}
    	\label{fig:Eig1}
    \end{figure}
        \begin{figure}[t!]
    	\centering
    	\includegraphics[width=0.95\textwidth]{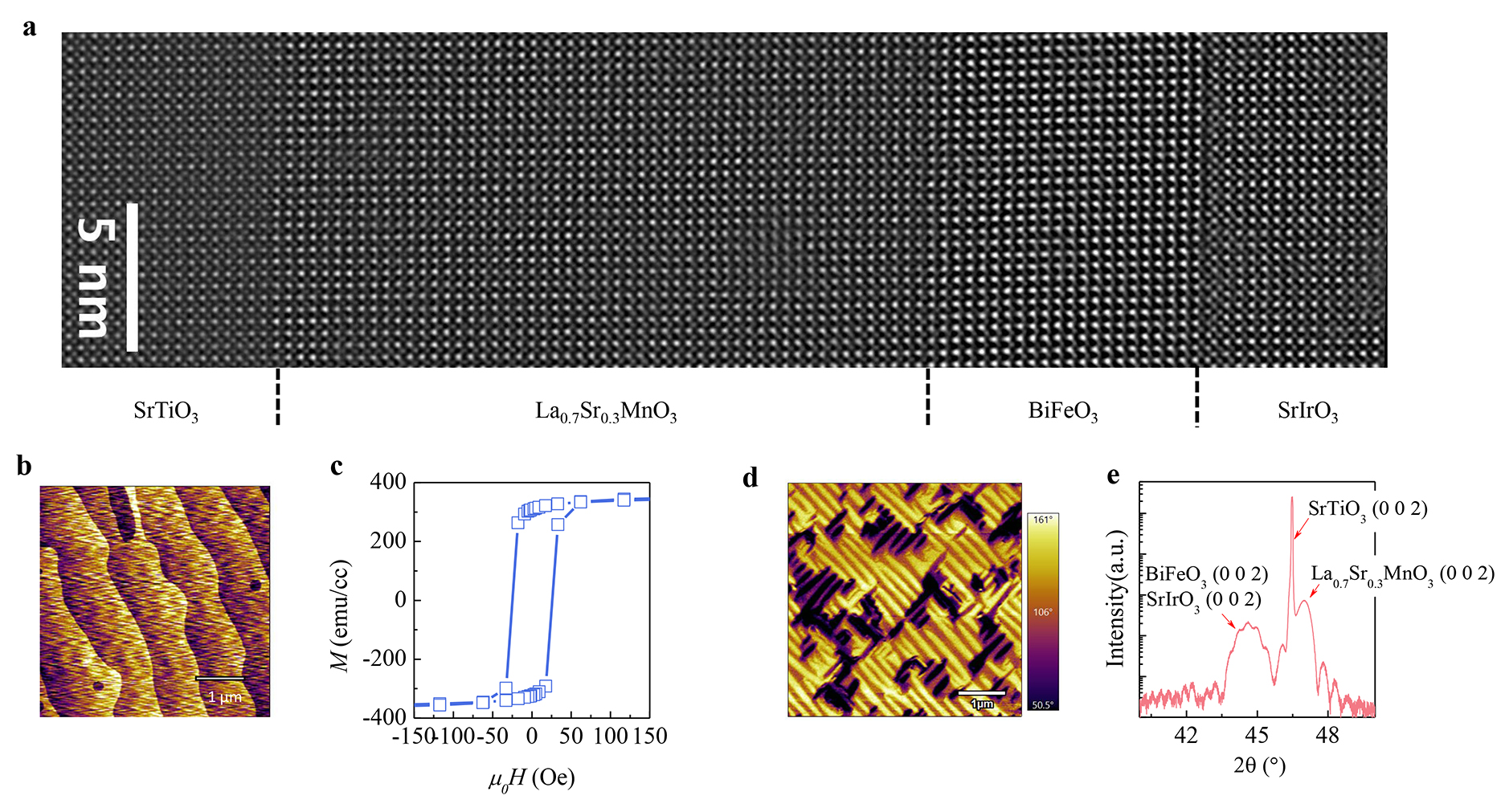}\caption{\textbf{Structure, magnetization, ferroelectric domain structure measurements for La$_{0.7}$Sr$_{0.3}$MnO$_3$/BiFeO$_3$/SrIrO$_3$ heterostructure. a,} HAADF image of La$_{0.7}$Sr$_{0.3}$MnO$_3$/BiFeO$_3$/SrIrO$_3$ tri-layer, displaying atomically sharp interfaces and high crystal quality. \textbf{b,} HF etched and thermally annealed SrTiO$_3$ substrate with atomic steps and terraces. \textbf{c,} Magnetization vs applied magnetic field (MH) measurement for La$_{0.7}$Sr$_{0.3}$MnO$_3$/BiFeO$_3$/SrIrO$_3$, showing a coercivity of $\sim$ 25 Oe and a saturation magnetization $\sim$ 320 emu/cc. \textbf{d,} In-plane piezoresponse microscopy (PFM) image of the BiFeO$_3$ layer. \textbf{e,} High resolution XRD 2$\theta$-$\omega$ scan of the tri-layer}
    	\label{fig:Eig2}
    \end{figure}

    \begin{figure}[t!]
    	\centering
    	\includegraphics[width=0.7\textwidth]{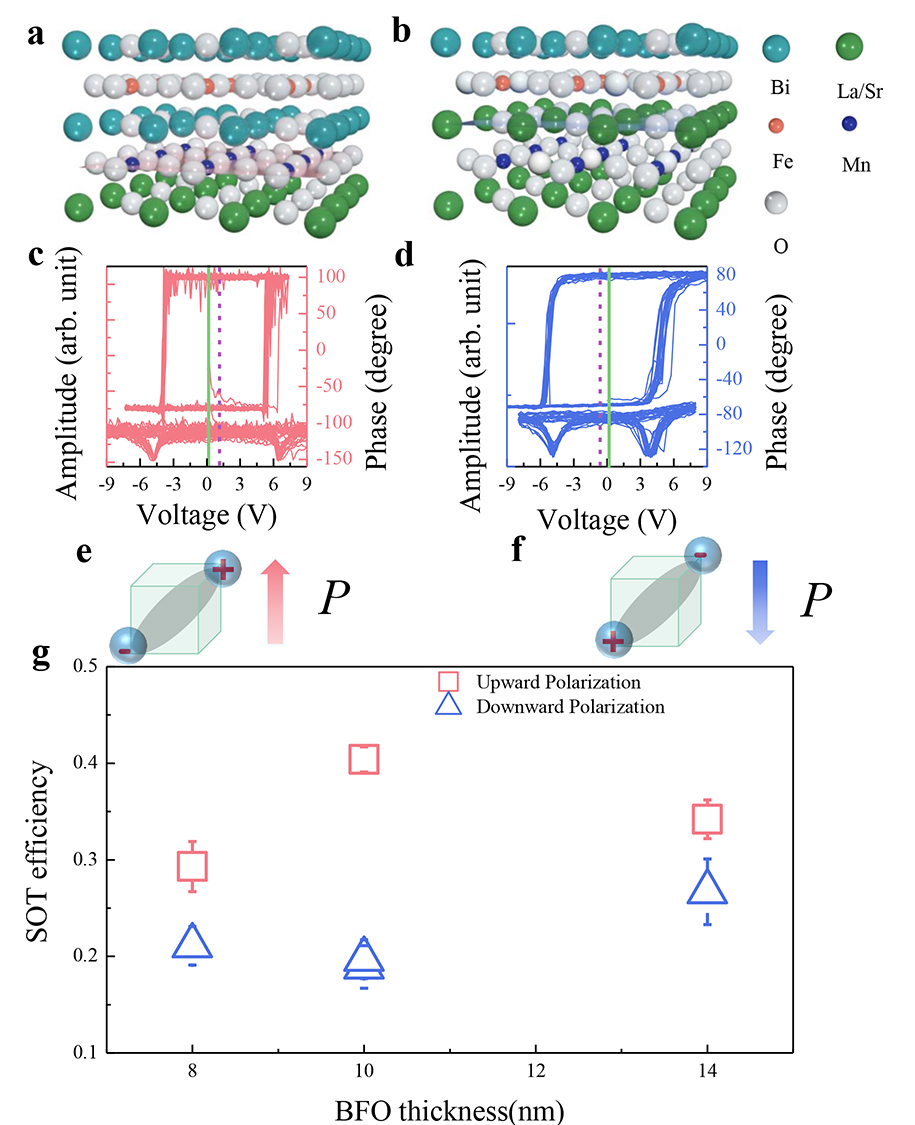}
    	\caption{\textbf{Ferroelectric control of spin transport enabled by the interface chemistry. a,} A schematic for La$_{0.7}$Sr$_{0.3}$MnO$_3$/BiFeO$_3$ atomic stacking, where La$_{0.7}$Sr$_{0.3}$O-MnO$_2$-BiO-FeO$_2$ is stacked at the interface. \textbf{b,} A schematic for La$_{0.7}$Sr$_{0.3}$MnO$_3$/BiFeO$_3$ atomic stacking, where MnO$_2$-La$_{0.7}$Sr$_{0.3}$O-FeO$_2$-BiO is stacked at the interface. \textbf{c,} Corresponding piezoresponse for La$_{0.7}$Sr$_{0.3}$O-MnO$_2$-BiO-FeO$_2$ stacking (La$_{0.7}$Sr$_{0.3}$MnO$_3$($12$ nm)/BiFeO$_3$($50$ nm)), with top and bottom being phase and amplitude respectively. The solid line denotes the shifting of the piezoresponse curves to zero volts, indicating an upward ferroelectric polarization for La$_{0.7}$Sr$_{0.3}$O-MnO$_2$-BiO-FeO$_2$ stacking. \textbf{d,} Corresponding piezoresponse for MnO$_2$-La$_{0.7}$Sr$_{0.3}$O-FeO$_2$-BiO stacking (La$_{0.7}$Sr$_{0.3}$MnO$_3$($12$ nm)/BiFeO$_3$($50$ nm)), with the top and bottom being phase and amplitude respectively. The solid line denotes the shifting of the piezoresponse curves to zero volts, indicating a downward ferroelectric polarization for MnO$_2$-La$_{0.7}$Sr$_{0.3}$O-FeO$_2$-BiO stacking. \textbf{e,f,} An artistic illustration for upward and downward ferroelectric polarization. \textbf{g,} Spin orbit torque efficiency for BiFeO$_3$ with different polarization states as a function of BiFeO$_3$ thicknesses. The red (blue) data points represent the spin-orbit torque efficiency for La$_{0.7}$Sr$_{0.3}$MnO$_3$/BiFeO$_3$/SrIrO$_3$, where BiFeO$_3$ has an upward (downward) polarization.}
    	\label{fig:Eig2}
    \end{figure}
    
    \begin{figure}[t!]
    	\centering
    	\includegraphics[width=1\textwidth]{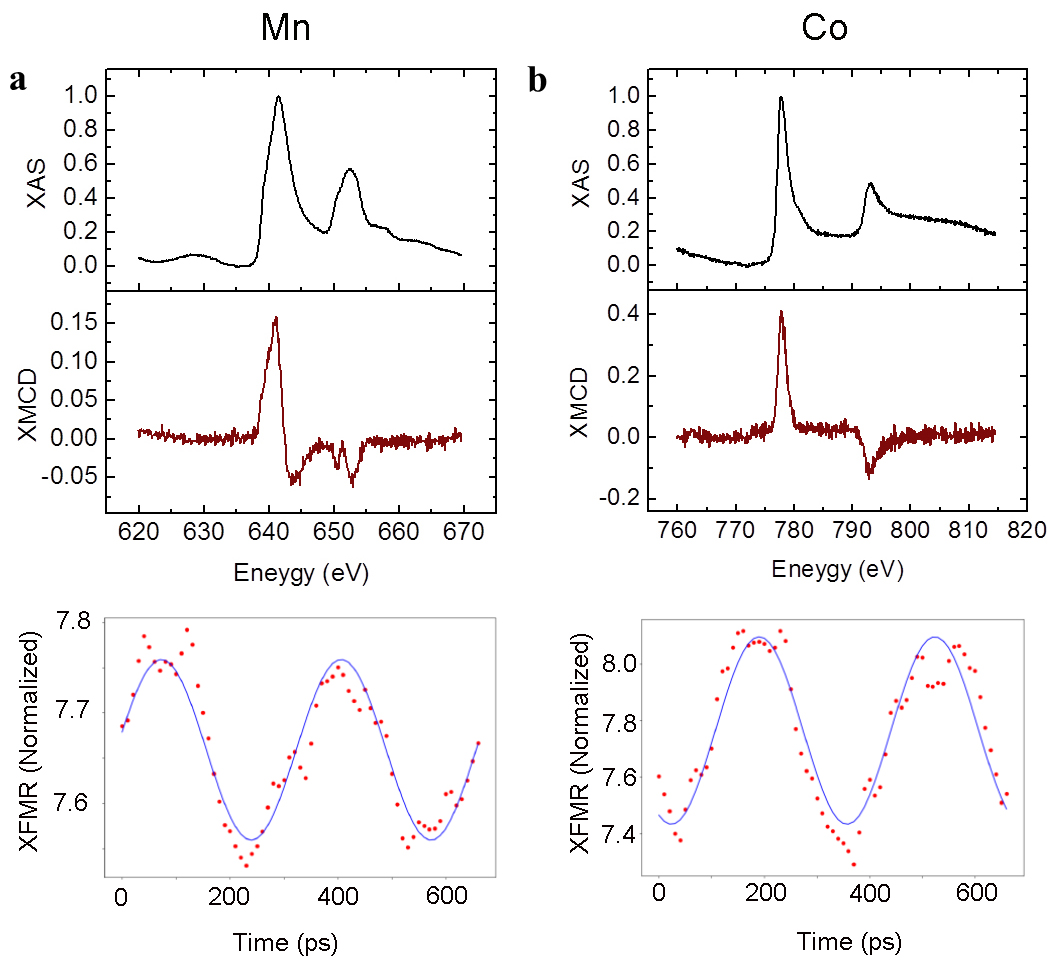}
    	\caption{\textbf{X-ray magnetic circular dichroism spectra for the X-ray Ferromagnetic resonance measurements in LSMO/BFO/Ti/CoFe Stacks. a, b,} X-ray absorption spectra (XAS), X-ray magnetic circular dichroism spectra (XMCD), and X-ray ferromagnetic resonance (XFMR) of Mn(a) and Co(b), which correspond to the magnetic element in LSMO and CoFe  respectively. The X-FMR measurements confirm the ac spin current transport through BFO. }
    	\label{fig:Eig3}
    \end{figure}
    \begin{figure}[t!]
    	\centering
    	\includegraphics[width=1\textwidth]{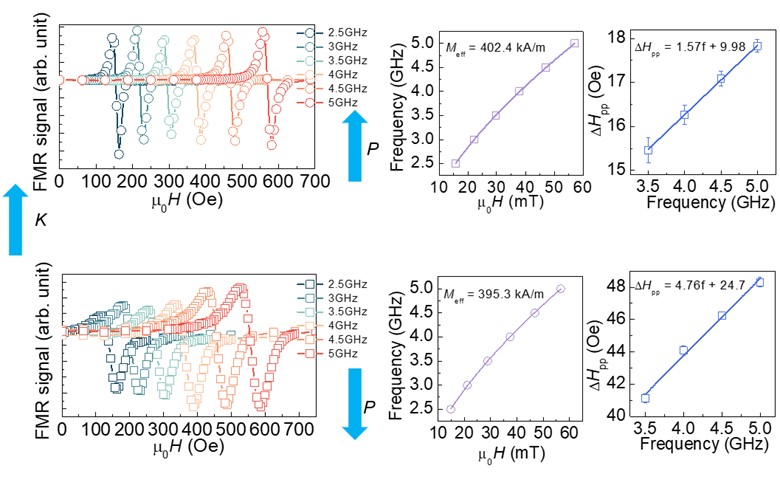}
    	\caption{\color{blue}\textbf{ FMR measurements for LSMO (20 nm)/BFO (10 nm)/SIO (10 nm) samples with different polarizations. a, b, }FMR spectrum for LSMO/BFO/SIO heterostructures with upward and downward ferroelectric polarizations respectively. Arrow K is a representation of spin propagation with respect to the polarization direction for FMR measurements. Resonance fields and frequencies are summarized in \textbf{c} and \textbf{d} for heterostructures with upward and downward respectively. Effective magnetization ~ 402 kA/m and 395 kA/m are determined with Kittel equation fitting. \textbf{e, f, } Peak-to-peak linewidth as a function of frequency is shown for heterostructures with upward and downward polarizations respectively.  Effective Gilbert damping constants ($\alpha$) of 0.00381 and 0.0116 are estimated for systems with upward and downward polarizations respectively} 
    	\label{fig:Efig6}
    \end{figure}

\begin{figure}[t!]
    	\centering
    	\includegraphics[width=1\textwidth]{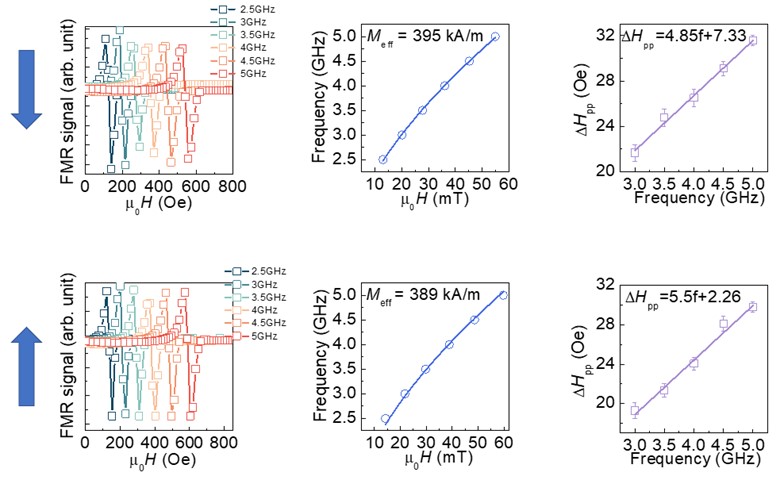}
    	\caption{{\color{blue}\textbf{FMR measurements for LSMO (20 nm)/BFO (40 nm) samples with different polarizations. a, b, }FMR spectrum for LSMO/BFO heterostructures with downward and upward ferroelectric polarizations respectively. Resonance fields and frequencies are summarized in \textbf{c} and \textbf{d} for heterostructures with downward and upward respectively. Effective magnetization ~ 395 kA/m and 389 kA/m are determined with Kittel equation fitting. \textbf{e, f, } Peak-to-peak linewidth as a function of frequency is shown for heterostructures with downward and upward polarizations respectively. Effective Gilbert damping constants of 0.01177 and 0.013335 are deduced for downward and upward polarizations respectively}} 
    	\label{fig:Efig6}
    \end{figure}

  \begin{figure}[t!]
    	\centering
    	\includegraphics[width=0.8\textwidth]{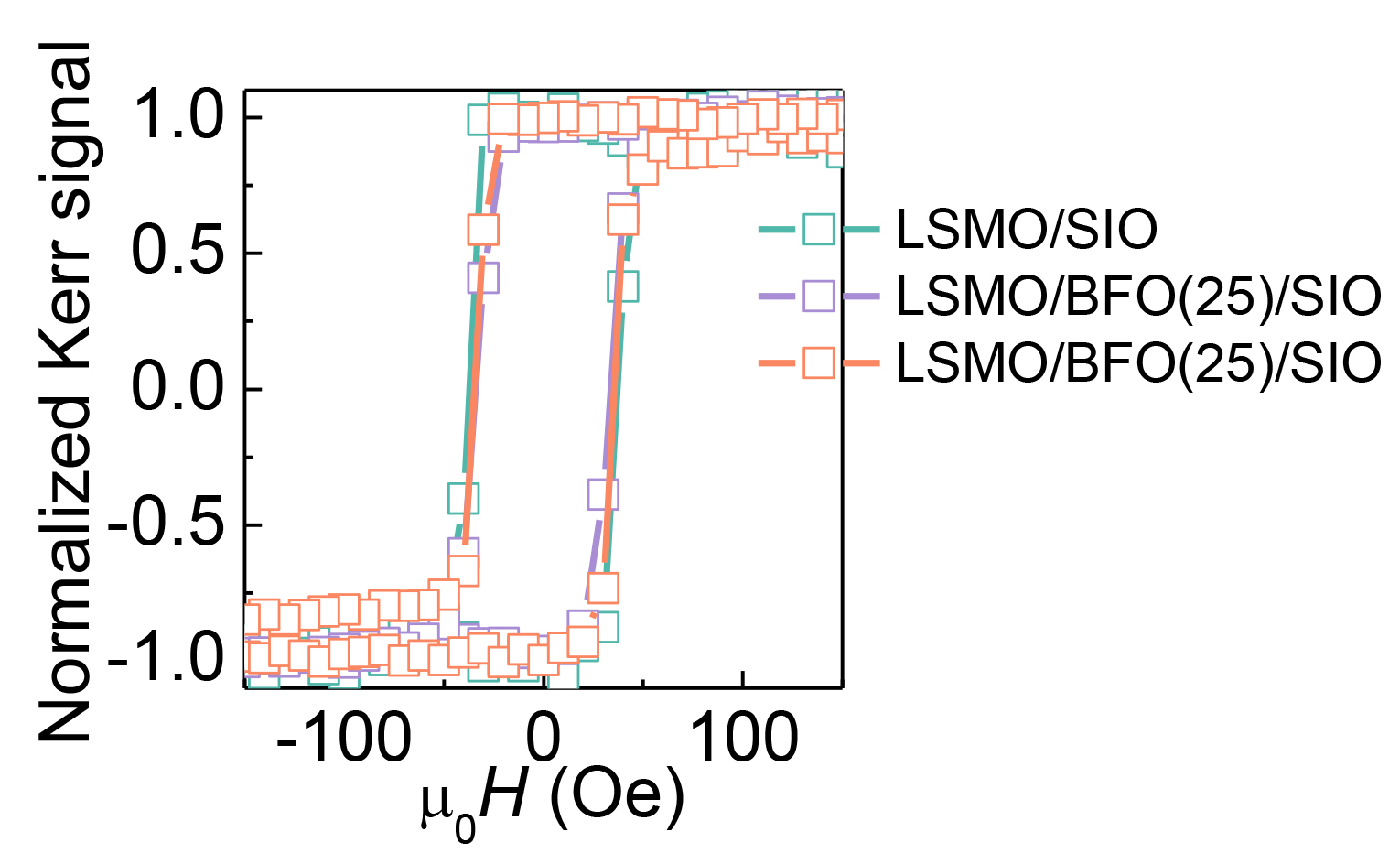}
    	\caption{\color{blue}\textbf{Exchange bias in BiFeO$_3$-based heterostructures.} MOKE measurements for LSMO (20 nm)/SIO (10 nm) and LSMO (20 nm)/BFO (25 nm)/SIO (10 nm) samples at room temperature. } 
    	\label{fig:Efig6}
    \end{figure}

    	\label{fig:Efig6}

   \begin{figure}[t!]
    	\centering
    	\includegraphics[width=0.7\textwidth]{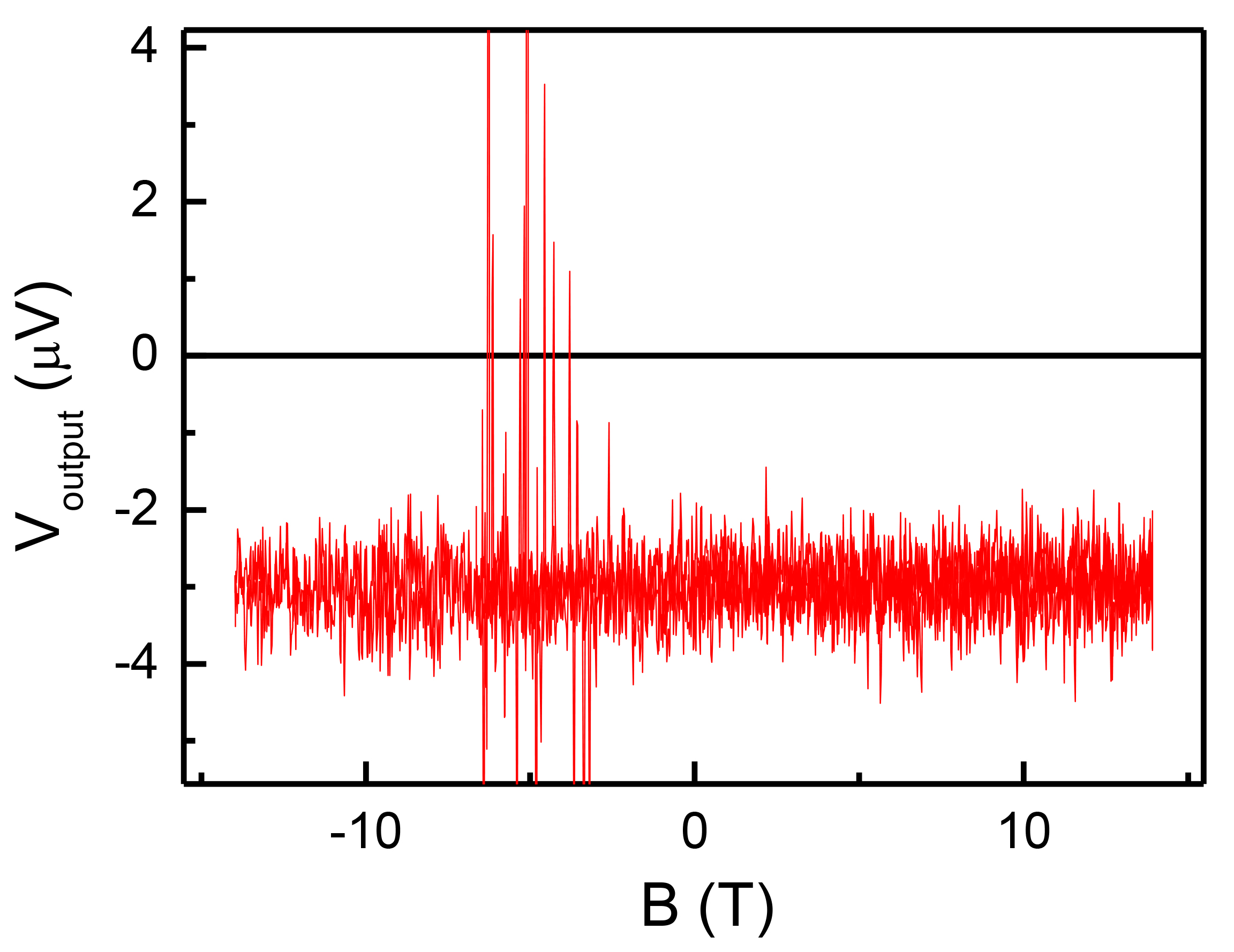}
    	\caption{{\color{blue}{\textbf{} dc non-local voltage as a function of applied magnetic field at room temperature. The applied dc current is 10 $\mu$A.}}} 
    	\label{fig:Efig6}
    \end{figure}

   \begin{figure}[t!]
    	\centering
    	\includegraphics[width=0.9\textwidth]{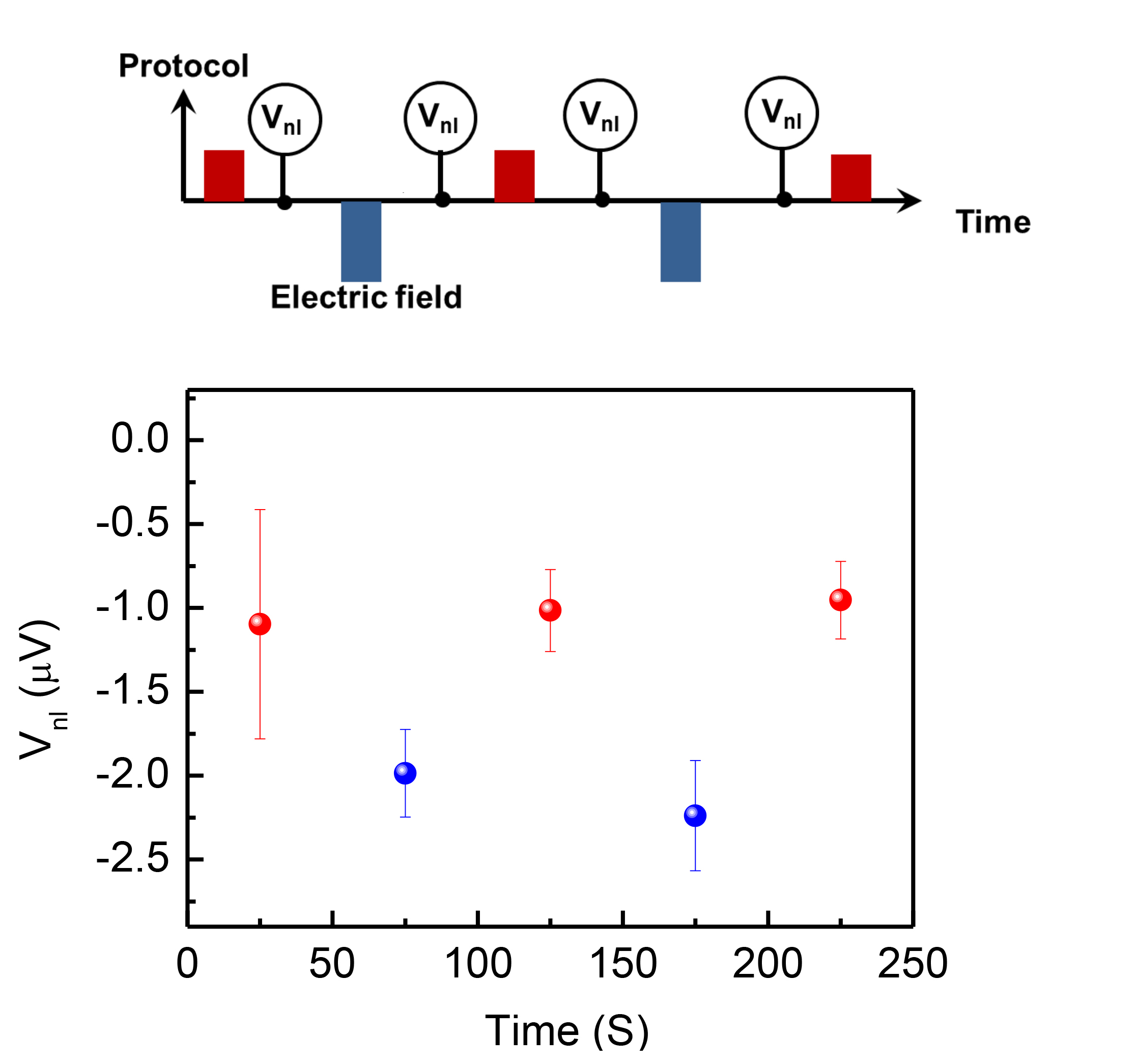}
    	\caption{{\color{blue}{\textbf{ Electric field controlled non-local spin transport measurements with dc current input.} A positive electric field of 150 kV/cm and a negative electric field of -300 kV/cm are applied alternately, with dc non-local voltages measured in between the electric fields as shown in the measurement protocol. The corresponding non-local voltage between each electric field is shown below. Each data point is averaged over a time interval of 50 s.}}} 
    	\label{fig:Efig6}
    \end{figure}

     \begin{figure}[t!]
    	\centering
    	\includegraphics[width=0.7\textwidth]{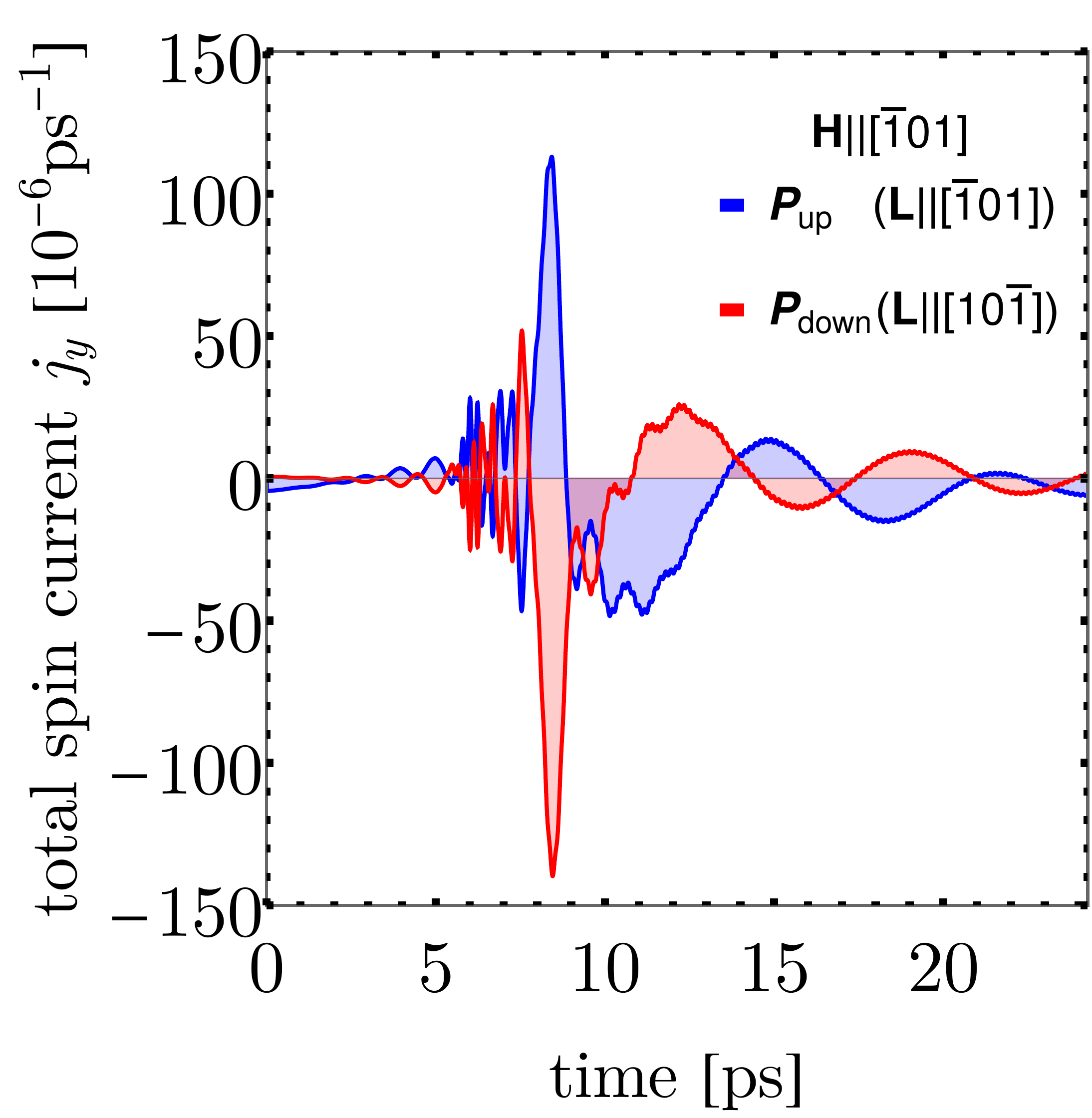}
    	\caption{\textbf{AFM spin current under {$\mathbf{L}$} switching.} Calculated $y$-component of the total spin current for up (blue) and down (red) polarization orientations. The applied magnetic field $\mathbf{H}$ to generate the excitation is set along [$\bar{1}01$]. The N\'{e}el vector in both polar configurations is different indicative of a possible $180^\circ$ switching scenario of $\mathbf{L}$. The ratio of the time-integrated \emph{dc} components is +0.978.} 
    	\label{fig:Efig6}
    \end{figure}

   \begin{figure}[t!]
    	\centering
    	\includegraphics[width=0.8\textwidth]{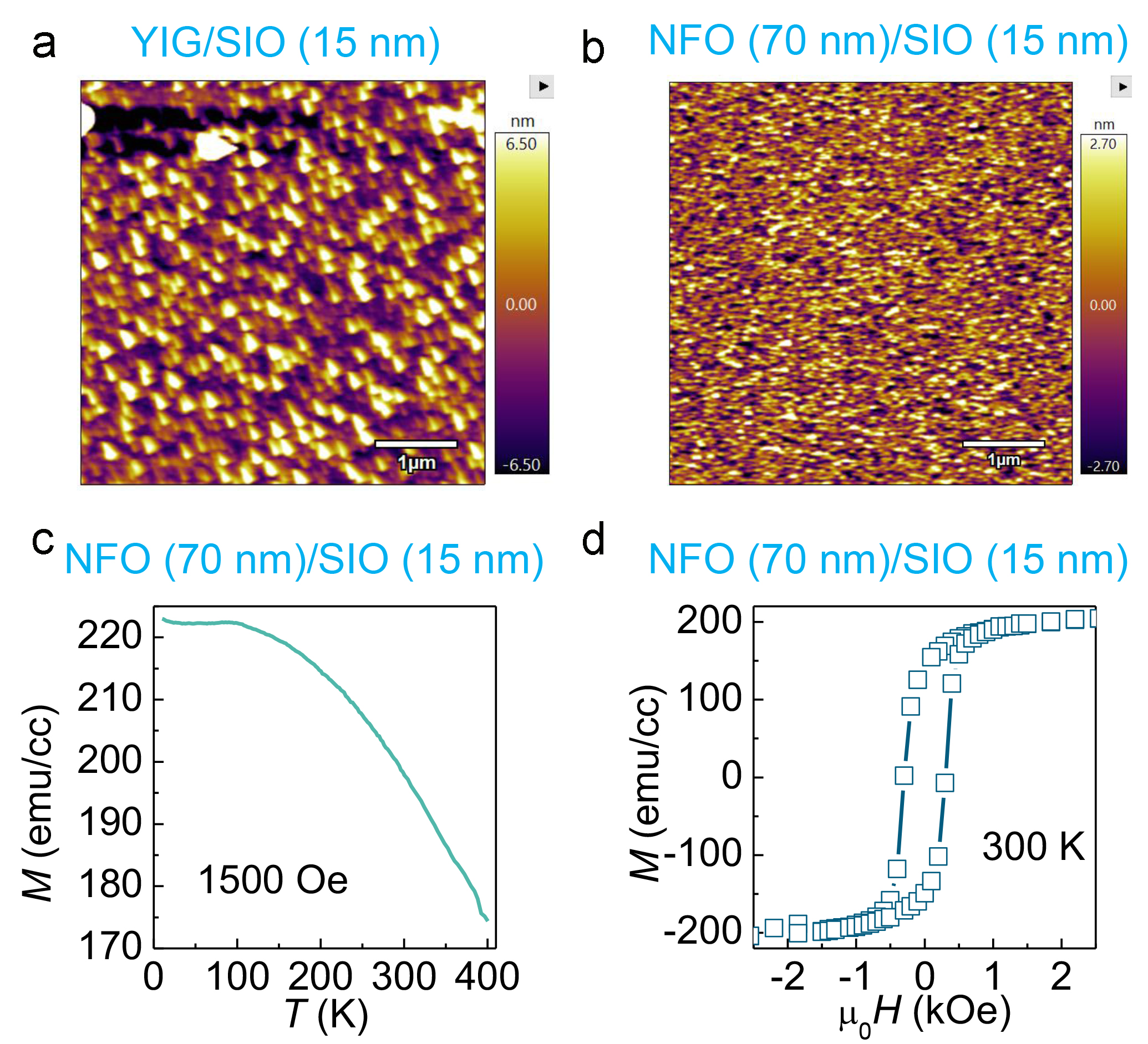}
    	\caption{\color{blue}\textbf{AFM and SQUID characterization. a, }An AFM image for SrIrO$_3$ (15 nm) grown on YIG/GGG structure. \textbf{b, }An AFM image for SrIrO$_3$ (15 nm) grown on NiFe$_2$O$_4$/STO structure. \textbf{c,} Magnetization as a function of temperature from 10 – 400 K, with a magnetic field of 1500 Oe applied in the plane. \textbf{d,} Magnetization as a function in-plane magnetic field at 300 K. } 
    	\label{fig:Efig6}
    \end{figure}

   \begin{figure}[t!]
    	\centering
    	\includegraphics[width=1\textwidth]{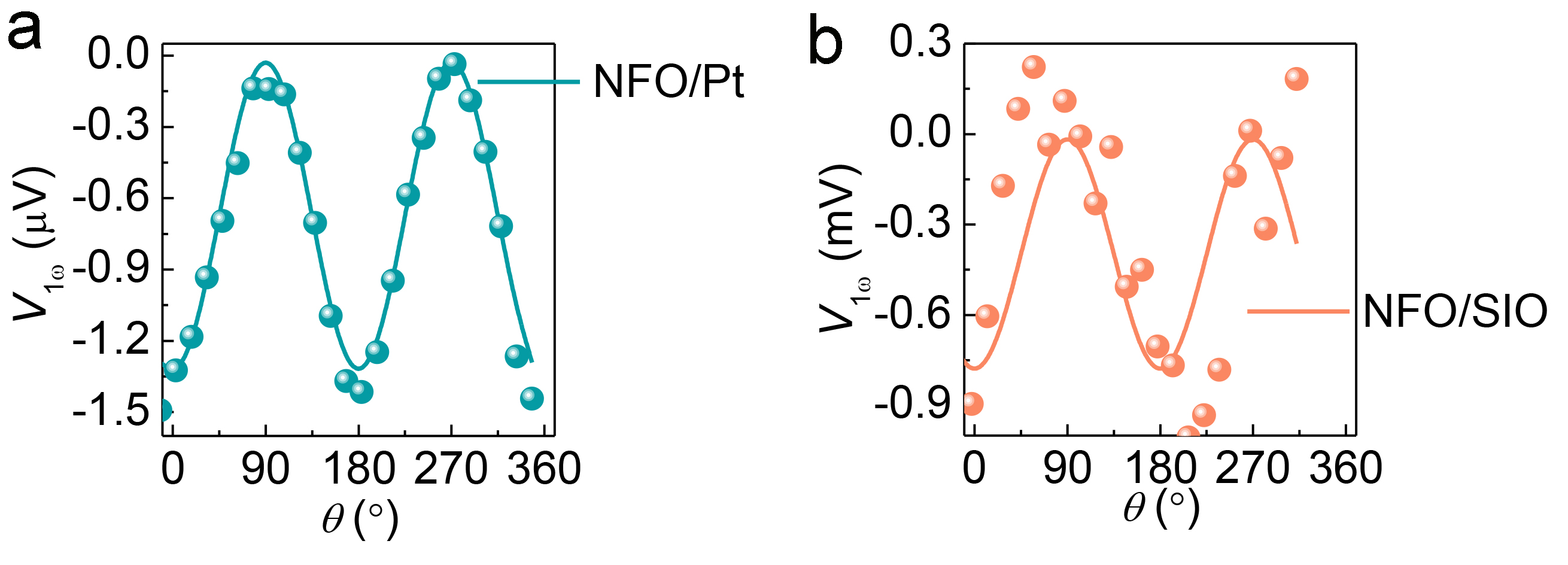}
    	\caption{\color{blue}\textbf{Non-local control experiments on NiFe$_2$O$_4$ (70 nm)/Pt (5 nm) and NiFe$_2$O$_4$ (70 nm)/SrIrO$_3$ (15 nm).} First harmonic voltage signal is displayed in \textbf{a} and \textbf{b} for NiFe$_2$O$_4$/Pt and NiFe$_2$O$_4$/SrIrO$_3$ respectively. An ac current with an amplitude of 100 $\mu$A and a frequency of 17 Hz was applied in a channel that is separated by 1 $\mu$m from the detection channel. } 
    	\label{fig:Efig6}
    \end{figure}

\end{document}


\maketitle
    
    \newpage

   {\color{blue} \section{Spin transport in a spin cycloid}
Since the spin cycloid propagates only inside the $xy$-plane along $[1\bar{1}0]$ direction, we can thus neglect the spin texture along $z$-direction. In the continuum limit, the magnetic property of BFO can thus be captured by the following free energy~\cite{Kadomtseva2006}
\begin{align}\label{continuous_free_energy}
F=\int{d{\bf{r}}^2}\{J[\nabla{\bf{L}}({\bf{r}})]^2+A[{\bf{m}}({\bf{r}})]^2+\mathcal{D}_K\hat{p}\cdot{\bf{L}}({\bf{r}})\times{\bf{m}}({\bf{r}})+\mathcal{D}_C{\bf{L}}({\bf{r}})\cdot[(\hat{p}\times{\bf{\nabla}})\times{\bf{L}}(\bf{r})]\},
\end{align}
where $\bf{L}({\bf{r}})$ (${\bf{m}}({\bf{r}})$) is the Ne\'el vector (total magnetization), $A=Ja^2$ with a the lattice constant, $\mathcal{D}_K=D_Ka^2$ and $\mathcal{D}_C=D_Ca$. 
The first and the second terms originate from the exchange interaction. The third term is the Dzyaloshinskii–Moriya interaction between nearest neighbor that parallel to the polarization, which results in a spin-canted, weak ferromagnetic structure~\cite{Albrecht2010}. The last term is the relativistic Lifshitz-like invariant between the next-nearest neighbor that leads to the magnetic cycloid in BFO. Note that the negligible anisotropy is not included in Eq.~\ref{continuous_free_energy}. Performing the differential equation $\delta{F}/\delta{\bf{m}}=0$ gives the ground state magnetization $m=D_K/J$, which is a high-order term compared to $L$. On the other hand, it has been shown that $D_K\ll D_C$~\cite{Albrecht2010}, the second and the third terms can thus be readily discarded. We can restore $\bf{m}$ later by using the relation that $\bf{m}$ is perpendicular to $\bf{L}$. Consequently, the ground state spin texture can be obtained by minimizing this total free energy by keeping only the first and the last term, which turns out to be a spin cycloid along $\hat{R}$ direction with $\hat{R}$ the unit vector inside the $xy$-plane that perpendicular to $\hat{p}$~\cite{Rahmedov2012}. In this case, the last term in Eq.~\ref{continuous_free_energy} can be intuitively viewed as an effect Dzyaloshinskii–Moriya interaction $\bf{D}$ perpendicular to both $\hat{p}$ and $\hat{R}$. We therefore obtain
\begin{align}\label{Neel}
{\bf{L}}({\bf{r}})=\begin{bmatrix}0,&\sin{(r_0D_C/J)},&\cos{(r_0D_C/J)}\end{bmatrix},
\end{align}
where $r_0={\bf{r}}\cdot{\hat{R}}$. Note that we have used a new coordinate system $\tilde{x}=\hat{d}={\bf{D}}/D$, $\tilde{y}=\hat{R}$, $\tilde{z}=\hat{p}$ for convenience. They are related to the lab coordinate system through 
\begin{align}\label{coor}
\begin{bmatrix}\tilde{x}\\ \tilde{y}\\ \tilde{z}\end{bmatrix}=\begin{bmatrix}d_x,&d_y,&d_z\\ {d_y}/{\sqrt{d_x^2+d_y^2}},&{-d_x}/{\sqrt{d_x^2+d_y^2}},&0\\ {d_xd_z}/{\sqrt{d_x^2+d_y^2}},&{d_yd_z}/{\sqrt{d_x^2+d_y^2}}, &-\sqrt{d_x^2+d_y^2}\end{bmatrix}\times\begin{bmatrix}\hat{x}\\ \hat{y}\\ \hat{z}\end{bmatrix}.
\end{align}
Moreover, the magnetization $\bf{m}$ that is perpendicular to $\bf{L}$ reads
\begin{align}\label{mm}
{\bf{m}}({\bf{r}})=\frac{D_K}{J}\begin{bmatrix}0,&\cos{(r_0D_C/J)},&-\sin{(r_0D_C/J)}\end{bmatrix}.
\end{align}

In the presence of this spin cycloid, the spin current transmit through BiFeO${}_3$ can be phenomenologically expressed as~\cite{lebrun2018}
\begin{align}\label{spin_current}
J_s=G_L\int{d{\bf{r}}^2}[{\bf{L}}({\bf{r}})\cdot{\bf{s}}]^2+G_m\int{d{\bf{r}}^2}[{\bf{m}}({\bf{r}})\cdot{\bf{s}}]^2,
\end{align}
where $G_{L}$ ($G_m$) is the phenomenological spin conductance of the N\'eel vector (total magnetization), and ${\bf{s}}=\hat{\bf{y}}$ represents the direction of spin accumulation inside the heavy metal. Substituting Eqs.~\ref{Neel} and ~\ref{mm} into Eq.~\ref{spin_current}, we finally find that the spin current transmit through BiFeO${}_3$ as a function of $\bf{D}$ can be expressed as
\begin{align}\label{int_spin_current}
\begin{split}
J_s({\bf{D}})&=G_L\int{d{\bf{r}}^2}[\frac{-d_x\sin{(r_0D_C/J)}}{\sqrt{d_x^2+d_y^2}}+\frac{d_yd_z\cos{(r_0D_C/J)}}{\sqrt{d_x^2+d_y^2}}]^2\\
&+\frac{G_mD_K}{L}\int{d{\bf{r}}^2}[\frac{-d_x\cos{(r_0D_C/J)}}{\sqrt{d_x^2+d_y^2}}-\frac{d_yd_z\sin{(r_0D_C/J)}}{\sqrt{d_x^2+d_y^2}}]^2,
\end{split}
\end{align}
where $d_x=\sin{\theta}\cos{\phi}$, $d_y=\sin{\theta}\sin{\phi}$, $d_z=\cos{\theta}$ with $\theta$ ($\phi$) the polar (azimuthal) angle of the effective Dzyaloshinskii–Moriya interaction $\bf{D}$. We have transformed the Ne\'el vector ${\bf{L}}({\bf{r}})$ and the magnetization ${\bf{m}}({\bf{r}})$ to the lab coordinate system by using Eq.~\ref{coor} in deriving Eq.~\ref{int_spin_current}.
The integral gives us
\begin{align}\label{spin_current_inal}
\begin{split}
J_s({\bf{D}})&=\frac{A(G_L+G_mD_K/L)}{2}\frac{d_x^2(1+d_z^2)}{d_x^2+d_y^2}\\
&=\frac{A(G_L+G_mD_K/L)}{2}\cos^2{\phi}(1+\cos^2{\theta}),

\end{split}
\end{align}
where $A$ is the area of the BFO in the $xy$-plane. The initial effective Dzyaloshinskii–Moriya interaction is along $[\bar{1}\bar{1}2]$ direction with the angles $\theta_0$ and $\phi_0$. However, in the presence of the built-in potential, this Dzyaloshinskii–Moriya interaction deviates from this initial direction by $\delta{\theta}$ and $\delta{\phi}$ as indicated in Fig~3 (d) in the maintext. Therefore $\theta=\theta_0+\delta{\theta}$ and $\phi=\phi_0+\delta{\phi}$. This deviation does NOT change under the reversal of the polariztion $\bf{P}$.
Only $\theta_0$ and $\phi_0$ flip their sign when $\bf{P}$ becomes $-\bf{P}$,
resulting in a spin current rectification.
To quantify this spin current rectification, we define a spin current difference parameter 
\begin{align}\label{spin_difference}
\xi=\frac{J_s(-{\bf{P}})-J_s({\bf{P}})}{J_s(-{\bf{P}})+J_s({\bf{P}})},
\end{align}
where characterizes the spin current rectification.

}
    
    \section{Antiferromagnetic spin wave calculations}

To simulate the ground state and subsequent spin wave dynamics of the antiferromagnetic (AFM) sublattices, we first relax the time-dependent Landau-Ginzburg-Devonshire equations, detailed in Ref. \cite{Shi2022}.
We concern ourselves with homogeneous states $(\mathbf{P}||\mathbf{A})||[111]$ and $(\mathbf{P}||\mathbf{A})||[\bar{1}\bar{1}\bar{1}]$ corresponding to the switched states in the experiment,
We define $\mathbf{P}$ as the electric polarization vector and $\mathbf{A}$ as the octahedral antiphase tilt field.
The resulting spontaneous order parameter values are $P_s = 0.94 \,\mathrm{C}/\mathrm{m}^2$  and $A_s = 13.4^\circ$.
To study spin waves, we propose a magnetic free energy density $F_\mathrm{mag}$ which is,
\begin{align}\label{eqn:mag_E}
    F_\mathrm{mag} &= D_e \left(\mathbf{L}^2 - \mathbf{m}^2\right) + 4 D_0 \textbf{A} \cdot \left( \textbf{L} \times \textbf{m}\right) + \sum\limits_{\eta = 1}^2 K_1 \left(\textbf{m}_\eta \cdot \hat{\textbf{P}}\right)^2\\ \nonumber
    &+ \sum\limits_{\eta = 1}^2 \left(K_1^c + a|\textbf{A}|^2\right)\left(m_{\eta,x}^2 m_{\eta,y}^2 m_{\eta,z}^2\right) + \frac{1}{2} \mathbf{m} \cdot \mathbf{H}_\mathrm{ext}.\\ \nonumber
    &+ A \left[(\nabla L_x)^2 + (\nabla L_y)^2 + (\nabla L_z)^2 \right]
\end{align}
where $\mathbf{m}_\eta$ are the sublattice magnetizations ($\eta = 1,2$). The normalized quantities are the N\'{e}el vector, $\mathbf{L} = \mathbf{m}_1 - \mathbf{m}_2$, and total magnetization $\mathbf{m} = \mathbf{m}_1 + \mathbf{m}_2$.
The variable $\hat{\textbf{P}}$ corresponds to the normalized director of $\textbf{P}$.
The free energy, $F_\mathrm{mag}$, contains contributions from the homogeneous AFM exchange \cite{Xu2019}, Dzyaloshinskii–Moriya interaction (DMI) \cite{Ederer2005, Rahmedov2012}, easy-plane magnetic and cubic anisotropy \cite{Weingart2012}, Zeeman interaction, and inhomogeneous exchange stiffness \cite{Agbelele2017}.
Next, we look for the ground states of the magnetization in the presence of the above fixed homogeneous orientations of $\mathbf{P}$ and $\mathbf{A}$.
To do this, we consider the Landau-Lifshitz-Bloch (LLB) equation \cite{Garanin2004} at zero Kelvin,
\begin{align}\label{eqn:LLB_1}
    \frac{d \mathbf{m}_\eta}{dt}&=-\gamma\left(\mathbf{m}_\eta\times\mathbf{H}_\eta\right)-\frac{\gamma\alpha}{1+\alpha^2}
    \mathbf{m}_\eta\times\left(\mathbf{m}_\eta\times\mathbf{H}_\eta\right) + \frac{\gamma\tilde\alpha_\parallel}{1+\alpha^2} m_\eta^2 \left[m_\eta^2 - 1\right]\mathbf{m}_\eta.
\end{align}
where $\mathbf{m}_\eta$ is the normalized magnetization of the $\eta^\mathrm{th}$ sublattice. The scalars $\gamma, \alpha$ and $\tilde\alpha_\parallel$ are the electron gyromagnetic ratio, phenomenological damping, and the longitudinal LLB damping constant respectively.
The latter term in Eq. \ref{eqn:LLB_1} provides a restoring force to constrain $|\mathbf{m}| + |\mathbf{L}| = 2$. The effective field is calculated via $\mathbf{H}_\eta = -(1/\mu_0 M_s) \delta F_\mathrm{mag} / \delta \mathbf{m}_\eta$.
Evolution of Eq. \ref{eqn:LLB_1} for nonzero $\alpha = 0.01$ leads to a 
competition of the nearest neighbor superexchange, DMI interaction, and single-ion easy-plane and cubic magnetocrystalline anisotropy terms respectively in Eq. \ref{eqn:mag_E} with $\mathbf{H}_\mathrm{ext} = 0$ and $A = 0$.
Thus, the breaking of the G-type antiferromagnetic order is seen as a weakly noncollinear AFM with $\textbf{P}$, $\textbf{m}$, and $\textbf{L}$ forming a right-handed coordinate system \cite{Ederer2005, Heron2014}.
The nonzero canting angle, $\phi = \cos^{-1}{\left(\mathbf{m}_1\cdot \mathbf{m}_2\right)}$, gives rise to a small non-vanishing $\mathbf{m}$ is determined by the ratio of $D_0/D_e$ which for BFO yields $\phi \approx 1.2^\circ$.
This noncollinearity leads to a magnetic moment of around $0.04 \,\mu$B/Fe in agreement with recent DFT observations \cite{Ederer2005, Wojdel2010, Dixit2015}.
The materials constants used for the $F_\mathrm{mag}$ are estimated from assorted references \cite{Weingart2012, Rahmedov2012, Dixit2015, Agbelele2017, Xu2019} and are listed in Table \ref{tab:coeff}.
Due to the weak cubic anisotropy \cite{Weingart2012}, there are six possible weak ferromagnetic and Néel vector domain orientations for each polar state (we have no reason to believe that our films support cycloidal order\cite{Burns2020} and look for homogeneous states). The list of $\mathbf{m}$ and $\mathbf{L}$ orientations for the ground states studied in this work are provided in Table \ref{tab:mag_states}.

\begin{table}[h!]
\centering
\begin{tabular}{c c c c}\toprule

         Symbols & Parameters & Values & Units \\
         \hline
         $D_e$ & AFM exchange coeff. & $6.82623\times 10^{-3}$ & eV/nm${}^3$\\
         $D_0$ & DMI coeff. & $4.93086\times 10^{-6}$ & eV/nm${}^3$ deg\\
         $K_1$ & easy-plane anisotropy & 0.0125781 & eV/nm${}^3$ \\
         $K_1^c$ & cubic SIA & 22 & $\mu$eV/nm${}^3$ \\
         $a$ & - & 1.462 & $\mu$eV/nm${}^3$ deg\\
         $A$ & inhomogeneous stiffness exchange & $3\times 10^{-7}$ & ergs/cm\\
         $\tilde\alpha_\parallel$ & LLB damping & 10 & unitless \\\bottomrule
\end{tabular}
\caption{Material coefficients for AFM spin wave model.}
\label{tab:coeff}
\end{table}

\begin{table}[h!]\centering
\begin{tabular}{c c c c c c c c}
\hline
& & & & & & & \\
 \textbf{P},\textbf{A}$||[111]$, $\mathbf{L} \simeq$ & $[\bar{1}10]$ & $[\bar{1}01]$ & $[0\bar{1}1]$ & $[1\bar{1}0]$ & $[10\bar{1}]$ & $[01\bar{1}]$ & \\ 
\,\,\,\,\,\,\,\,\,\,\,\,\,\,\,\,\,\,\,\,\,\,$\mathbf{m} \simeq$ & $[\bar{1}\bar{1}2]$ & $[1\bar{2}1]$ & $[2\bar{1}\bar{1}]$ & $[11\bar{2}]$ & $[\bar{1}2\bar{1}]$ & $[\bar{2}11]$ & \\ 
 \textbf{P},\textbf{A}$||[\bar{1}\bar{1}\bar{1}]$, $\mathbf{L} \simeq$ & $[\bar{1}10]$ & $[\bar{1}01]$ & $[0\bar{1}1]$ & $[1\bar{1}0]$ & $[10\bar{1}]$ & $[01\bar{1}]$ & \\  
\,\,\,\,\,\,\,\,\,\,\,\,\,\,\,\,\,\,\,\,\,\,$\mathbf{m} \simeq$  & $[11\bar{2}]$ & $[\bar{1}2\bar{1}]$ & $[\bar{2}11]$ & $[\bar{1}\bar{1}2]$ & $[1\bar{2}1]$ & $[2\bar{1}\bar{1}]$ &\\ 
\hline
\end{tabular}
\caption{\label{tab:mag_states}%
Sixfold degenerate magnetic ground states for up and down polar orientations. Note that these listed directions do not contain normalization prefactors. Due to the DMI coupling, full reversal of $\mathbf{P}$ displays full reversal of $\mathbf{m}$.} 
\end{table}

In the Fig.~1(c) in the main text, we demonstrate spin waves in the ground states with $\mathbf{L} = [\bar{1}01]/\sqrt{2}$ for $\mathbf{P}||[111]$ and $\mathbf{P}||[\bar{1}\bar{1}\bar{1}]$ (see Table \ref{tab:mag_states}).
To do this, we introduce an external field via the Zeeman term where $\mathbf{H}_\mathrm{ext}$ takes the form of a $\mathbf{H}_0 \exp{[-p_0(z-z_0)^2]}\cdot\mathrm{sinc}[k_0 (z-z_0)]\cdot\mathrm{sinc}[\omega_0 (t-t_0)]$ pulse in space and time with an amplitude of $H_0 = |\mathbf{H}_0| = 0.42$ Oe with $k_0 = 10$ nm${}^{-1}$, $z_0 = 30$ nm, $p_0 = 0.16$ nm${}^{-2}$, $t_0 = 1$ ps, and $\omega_0 = 1$ THz.
We emphasize that the observed differences in the components of the \emph{dc} current in the main text, is independent of the the exact method excitation. Similar torques/initial conditions may be obtained via applied polarized spin current. However, due to the complexity of the spin-current induced torques in AFMs \cite{Baltz2018, Xue2022}, we considered the simplest model with just an applied external magnetic field.

Equations \ref{eqn:mag_E} and \ref{eqn:LLB_1} are cast into weak forms sufficient for finite element analysis and solved in the Ferret module \cite{Mangeri2017, ferretlink} which is part of the open-source Multiphysics Object Oriented Simulation Environment (MOOSE) framework \cite{Permann2020}.
We consider periodic boundary conditions on $\mathbf{m}_\eta$ along the $\hat{\mathbf{k}}$ direction and a computational volume of $5\times5\times60$ nm. We use a time step of $10$ fs and find no differences in the dynamics when the time step is reduced.
Addtionally, we find no variation in the wave behavior or the constraint $|\mathbf{m}|+|\mathbf{L}| = 2$ when the phenomenological LLB damping parameter $\tilde\alpha_\parallel$ is increased.

\section{Discussion of spin dynamics}

In this material the direction of the chiral interaction corresponding to the antisymmetric exchange coupling (DMI) is given by the octahedral antiphase tilt vector $\mathbf{A}$ \cite{Rahmedov2012}. This chiral interaction favors one sense of rotation with respect to the other. Therefore, by reversing the direction of $\mathbf{A}$, one automatically switches the preferred chirality of the spin waves. In simpler terms, notice that by switching the direction of $\mathbf{A}$ in the following component of the energy, one changes the contribution given by perturbations of the ground state.
In the linearized limit of Eq \ref{eqn:LLB_1}, the DMI energy

$$F_\mathrm{DMI,P} = 4 D_0 \mathbf{A}\cdot \left[\left(\mathbf{L}+\delta \mathbf{L}\right)\times \left(\mathbf{m}+\delta\mathbf{m} \right)\right]$$
which under switching of $\mathbf{A} \to -\mathbf{A}$ and $\mathbf{m}\to -\mathbf{m}$ leads to different torques on the sublattices which producing a change in chirality.

\begin{figure*}[hp!]
    \centering
    \includegraphics[width=0.8\linewidth]{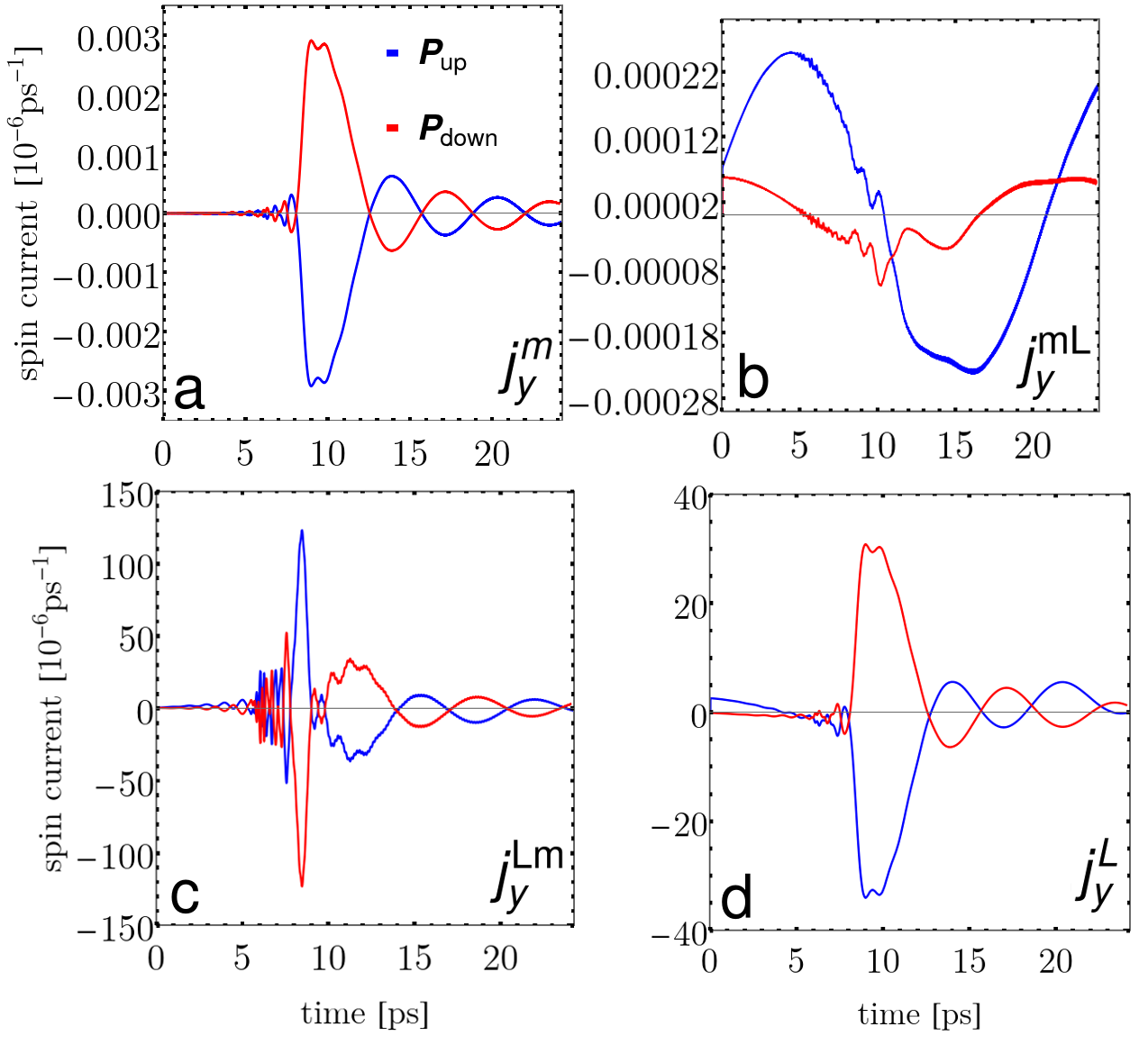}
    \caption{\textbf{Decomposition of AFM spin current} y-components of the spin currents proportional to (a) $\mathbf{m}\times\dot{\mathbf{m}}$ (b) $\mathbf{m}\times\dot{\mathbf{L}}$ (c) $\mathbf{L}\times\dot{\mathbf{m}}$ (d) $\mathbf{L}\times\dot{\mathbf{L}}$ after the $\mathbf{H}||[\bar{1}01]$ excitation at $\omega_0 = 1$ THz with an amplitude of $H_0 = 0.42$ Oe of a spin wave along $\mathbf{k}||[001]$.}
    \label{fig:decomp}
\end{figure*}

In this work, we consider the pumped spin currents in AFMs proposed in Ref. \cite{Cheng2014}. Due to the precession of both $\mathbf{m}$ and $\mathbf{L}$, the total spin current is comprised of four terms,
\begin{align}\label{eqn:j_spin}
    \mathbf{j} &= \mathbf{j}^\mathbf{m} + \mathbf{j}^{\mathbf{mL}} + \mathbf{j}^{\mathbf{Lm}} + \mathbf{j}^{\mathbf{L}}. 
\end{align}
with $\mathbf{j}^\mathbf{m} \propto \mathbf{m}\times\dot{\mathbf{m}}$, $\mathbf{j}^\mathbf{mL} \propto \mathbf{m}\times\dot{\mathbf{L}}$, $\mathbf{j}^\mathbf{Lm} \propto \mathbf{L}\times\dot{\mathbf{m}}$, and $\mathbf{j}^\mathbf{L} \propto \mathbf{L}\times\dot{\mathbf{L}}$. Here we neglect prefactors due to the spin-mixing conductance and only concern ourselves with the spin currents dependent on the normalized quantities.
As such, the units of our spin current are presented in ps${}^{-1}$.
Due to the different left or right-handed precession of spins in each ``up" or ``down" domain, the spin current vector components ($\mathbf{j}^\mathbf{Lm}, \mathbf{j}^\mathbf{m},\mathbf{j}^\mathbf{L}, \mathbf{j}^\mathbf{mL}$) exhibit phase differences.
This phase difference is shown in Fig.~1(c) in the main text and Fig.~\ref{fig:decomp} for each contribution to $\mathbf{j}$.
From Fig.~\ref{fig:decomp}, an important observation is that the magnitudes of the spin current contributions obey the following relationship,
$|\mathbf{j}^\mathbf{Lm}| > |\mathbf{j}^\mathbf{L}| >> |\mathbf{j}^\mathbf{m}| \approx |\mathbf{j}^\mathbf{mL}|$ reflecting the fact that $|\mathbf{m}| << |\mathbf{L}|$ in this material and notably, that the phase differences in these quantities between the $\mathbf{P}$ ``up" and $\mathbf{P}$ ``down" orientations arises from the above arguments for the DMI-induced torques.
Similar relationships are obtained for the other spatial components of $\mathbf{j}$.

\begin{figure*}[hp!]
    \centering
    \includegraphics[width=0.8\linewidth]{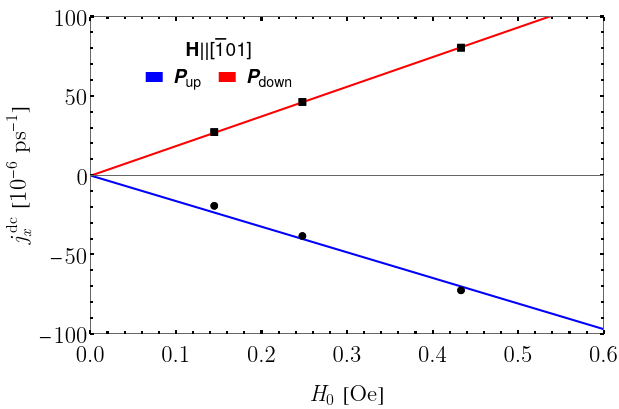}
    \caption{\textbf{Linear spin wave regime}: Calculation of $x$-component of total spin currents for different polarization orientations shown in Fig.~1(c) of the main text. Similar linear behavior is observed in the $y$ and $z$ components as well.}
    \label{fig:lin_spin}
\end{figure*}

We time integrate over the excitation window to find the \emph{dc} components of the spin current.
To verify that our calculations are in the linear regime, we change the amplitude of the applied magnetic field from $H_0 = 0.14$ to $0.42$ Oe and verify that the calculated dc components of the spin current for both $\mathbf{P}$ orientations obey a linear relationship as shown in Fig.~\ref{fig:lin_spin}.

    \section{Spin circuit model of magnon-mediated magnetoelectric spin-orbit logic} 
\begin{figure*}
    \centering
    \includegraphics[width=1.05\linewidth]{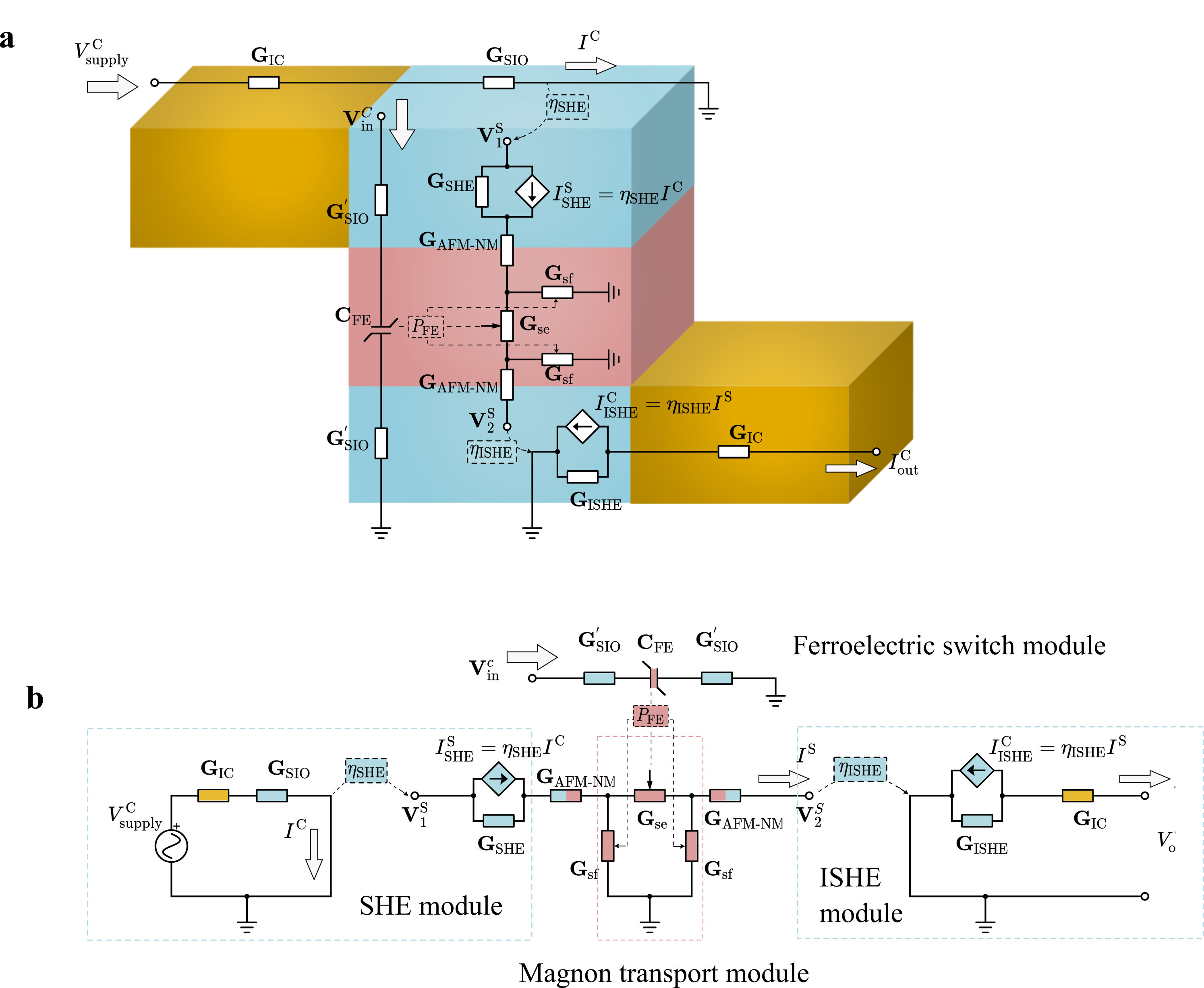}
    \caption{\textbf{Spin circuit model of the compact multiferroic logic device. a,} Circuit model overlaying the device. In the device layout, the yellow blocks represent metal interconnects, the blue blocks are strontium iridate (SIO), and the pink block is the bismuth ferrite (BFO). The physical interpretations and values of the circuit element symbols are listed in Table \ref{tab:element}. 
    \textbf{b,} Circuit model replicating that in Fig.~\ref{fig:circuit1}(a), with emphasis of each module.  
    modularized circuit schematics.  
    	The top module is the ferroelectric switch module, where the input voltage $V_\mathrm{in}^c$ switches the direction of the  ferroelectric polarization $P_\mathrm{FE}$ in the magnetoelectric layer.  
    	Such polarization change also affect the spin conductance values $G_\mathrm{se}$ and $G_\mathrm{sf}$ of the ME layer. 
    	The leftmost and rightmost modules represent the spin Hall effect (SHE) process in the top SOC layer (injector) and inverse spin Hall effect (ISHE) processe in the bottom SOC layer (detector), respectively. 
    	The current-controlled spin current source $I^S_\mathrm{SHE}$ in the SHE module depends on the output charge current of in the top SOC layer. 
    	Similarly, the current-controlled charge current source $I^C_\mathrm{ISHE}$ in the ISHE module depends on the output spin current of the bottom SOC layer. 
    	The parameters $\eta_\textrm{SHE}$ and $\eta_\textrm{ISHE}$ represent the charge-to-spin and spin-to-charge current conversion rates, respectively.
    	The module in the middle describes the magnon transport process in the ME layer, which is connected to the SHE and ISHE modules through the spin conductance at the interfaces between ME and SOC layers, denoted as $G_\mathrm{AFM-NM}$. }
    \label{fig:circuit1}
    \end{figure*}

    Circuit representations of spintronic devices have been a convenient tool to analyze the electric signal response of these devices, as they can be readily applied in conjunction with conventional circuit simulators such as SPICE.  
    We have constructed the equivalent circuit model of the chiral magnon mediated magnetoelectric spin orbit logic device based off the vector spin circuit theory \cite{Brataas2006, Manipatruni2012, manipatruni2019}, and adjusted the circuit formulation to better describe our system. 
    Fig.~\ref{fig:circuit1}(a) shows the spin circuit model overlaying the device structure.
    To investigate each step in the energy transduction process, we adopt the modular approach. 
    As shown in Fig.~\ref{fig:circuit1}(b), there are four main modules: ferroelectric switch module, spin Hall effect (SHE) module, magnon transport module, and inverse spin Hall effect (ISHE) module.
    The magnon transport module is connected to the SHE and ISHE modules through the interface spin conductance between the magnetoelectric (ME) layer and the spin-orbit-coupling (SOC) layer. 
    The rest of this section elaborates the derivation of each module and calculates the values of the circuit element parameters. 
    
   \subsection{Ferroelectric switch module}\label{subsec:FE}
    This module is composed of the electrical conductance $G^{'}_\mathrm{SIO}$ of the SOC layer and the capacitance $C_\mathrm{FE}$ of the ME layer.
    The subscript $SIO$ is used in the conductance term to directly indicate the SOC material we used in the design.
    When a switch voltage $V_\mathrm{in}^c$ is applied from the top SIO layer, an electric current is excited and flowing through the stacks of SIO and BFO in the vertical direction. 
    In the SIO layers, the current flow can be simply described by Ohm's law, resulting in an electrical conductance of $G^{'}_\mathrm{SIO} = \sigma_\mathrm{SIO}A/t$, where $A$ is the cross sectional area of the vertical current flow, i.e. $A = l_{\scaleto{\mathrm{SIO}}{4pt}} \times w_{\scaleto{\mathrm{SIO}}{4pt}}$, and $t$ is the dimension along the current flow direction, which is the film thickness $t_{\scaleto{\mathrm{SIO}}{4pt}}$.
    Similarly, the ME layer (which is BFO material in our setup) is equivalently an electric capacitor, due to its electrical insulating property. 
    We should note that the BFO capacitance $C_\mathrm{FE}$ exhibits ferroelectric properties, as indicated by its notation, which is nonlinear and dependent on the input voltage magnitude. 
    When the input voltage $V_\mathrm{in}^c$ changes the sign, the ferroelectric polarization $P_\mathrm{FE}$ in the ME layer is switched from $-z$ to $+z$ or vice versa.
    This polarity change will affect the spin conductance values $G_\mathrm{se}$ and $G_\mathrm{sf}$ in the magnon transport module (introduced in Section \ref{subsec:Magnon}), and this is a circuit representation of electrically controlled magnon transport.
    
    In this work, we focus on the magnon behavior and thus have omitted the ferroelectric switch module in the circuit simulations. 
    Our future work includes the precise calculation of $C_\mathrm{FE}$ with phase-field models \cite{Chen2008} described by the time-dependent Landau-Ginzburg-Devonshire equations. 
    Once the polarization $P$ is successfully simulated, the ferroelectric capacitance can be calculated as
    \begin{equation}
    C_{\rm FE} = \frac{\partial^2 U}{\partial Q_F^2}
    \label{free}
    \end{equation}
    where the energy density $U = f_{\rm Land} = \frac{1}{2}\alpha P^2 + \frac{1}{4}\beta P^4 + \frac{1}{6}\gamma P^6$ and $Q_F = P \times A$, with $A$ being the cross sectional area.
    
    With the SIO physical parameters summarized in Table \ref{tab:SIO}, we have estimated the values of $G^{'}_\mathrm{SIO}$ and $C_\mathrm{FE}$ shown in Table \ref{tab:element}.
    
    \subsection{SHE module}\label{subsec:SHE}
    As a horizontal supply voltage $V_\mathrm{supply}^c$ is applied to the top metal interconnect (yellow blocks in Fig.~\ref{fig:circuit1}(a)), a charge current $I^C$ is induced and flow into the SIO layer (blue blocks in Fig.~\ref{fig:circuit1}(a)), following Ohm's law. 
    We apply similar conductance formula as in Section \ref{subsec:FE}, with the cross section of the current flow being $A=w_{\scaleto{\mathrm{SIO}}{4pt}} t_{\scaleto{\mathrm{SIO}}{4pt}}$ and the longitudinal dimension being $l_{\scaleto{\mathrm{SIO}}{4pt}}$.
    Refer to Table \ref{tab:element} for the values of $G_\mathrm{SIO}$ and $G_\mathrm{IC}$.
    
\begin{figure*}
    \centering
    \includegraphics[width=0.8\linewidth]{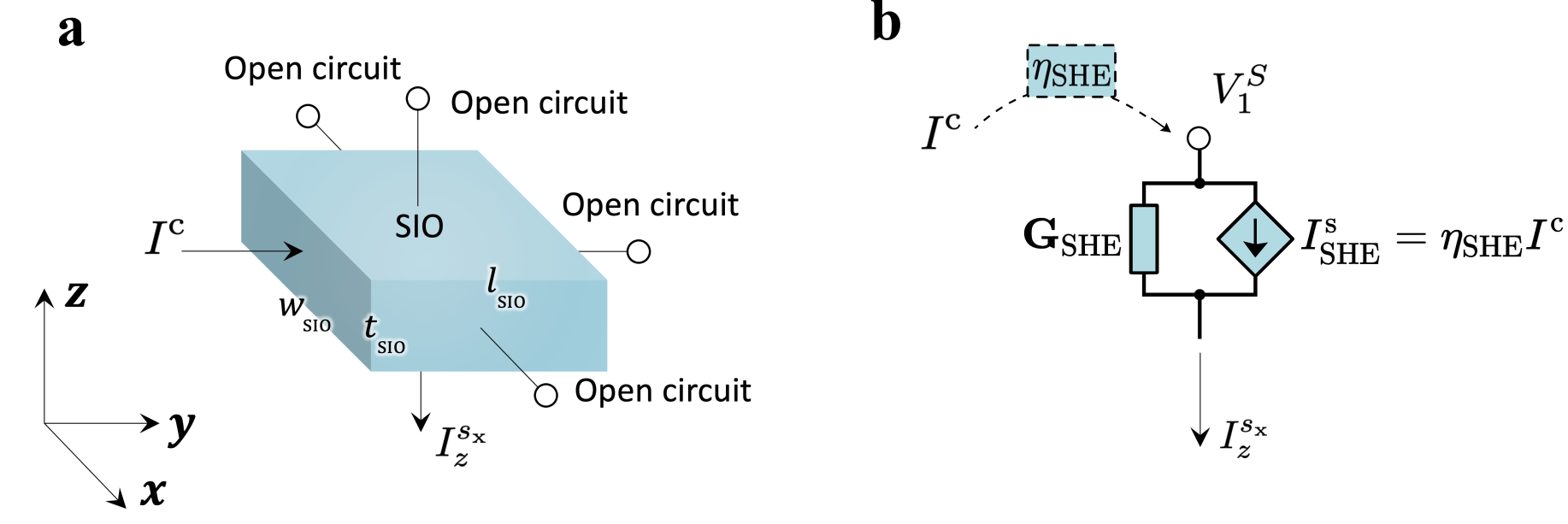}
    \caption{\textbf{Spin Hall effect circuit module. a,} The SIO block with dimensions of $l_{\scaleto{\mathrm{SIO}}{4pt}} \times w_{\scaleto{\mathrm{SIO}}{4pt}} \times t_{\scaleto{\mathrm{SIO}}{4pt}}$. The charge current is fed into the SIO material from the left terminal along the $\hat y$ direction, and excites a spin current along the $- \hat z$ direction flowing out of the bottom terminal. Note that the ``x'' in the $I^{s_x}_z$ notation indicates the direction of the spin polarization, instead of the spin transport direction (indicated by the subscript ``z''). All the other terminals are open-circuited. \textbf{b,} Equivalent spin Hall effect circuit module, represented by a current-controlled current source $I^s_\text{SHE}$, with an internal conductance $G_\text{SHE}$ \cite{Hong2016}. The variable $\eta_\text{SHE}$ is the charge-to-spin current conversion rate.  }
    \label{fig:SHmod}
    \end{figure*}

    Through SHE, polarized spins are generated  from the charge current in the top SIO layer. 
    To analyze this process, we follow the general framework proposed by Hong et al. \cite{Hong2016}. 
    Our preknowledge of the device setup allows us to capture the essential charge-to-spin conversion properties with simplified boundary conditions of the SIO material.
    Therefore, only two ports are active out of the six, as shown in Fig.~\ref{fig:SHmod}(a). 
    A charge current flows into the SIO material from the left surface and the spin current is generated and injected into the next layer of material from the bottom surface. 
    This simplified scenario can be described by a current-controlled current source $I^S_\text{SHE}$, with an internal conductance $G_\text{SHE}$ \cite{Hong2016}. 
    The spin current $I^S_\text{SHE}$ is dependent on the charge current $I^\text{C}$ with the relation $I^S_\text{SHE} = \eta_\text{SHE} I^\text{C}$.
    The factor $\eta_\text{SHE}$ is defined as
    \begin{equation}
    \eta_{\scaleto{\mathrm{SHE}}{4pt}} = \frac{l}{t} \frac {\theta_\text{SHE}}{1+ \frac{\lambda \theta ^2 _\text{SHE}} {t} \text{tanh} \frac{t}{\lambda} } \left( 1- \text{sech} \frac{t}{\lambda} \right) 
    \label{eq:etaSH}
    \end{equation}
    The internal conductance of the current source is obtained by dividing the short-circuit current by the open-circuit voltage, given by
    \begin{equation}
    G_\mathrm{SHE} = \sigma \frac{wl}{t} \frac{2 \theta^2_\mathrm{SHE} + \frac{t}{\lambda} \mathrm{coth \frac{t}{2 \lambda}}}{1+ \frac{\lambda \theta ^2 _\text{SHE}} {t} \text{tanh} \frac{t}{\lambda} } \left( 1- \text{sech} \frac{t}{\lambda} \right) 
    \label{eq:GSH}
    \end{equation}
    In Equation (\ref{eq:etaSH}) and (\ref{eq:GSH}), $\theta_\mathrm{SHE}$ is the spin Hall angle, $\sigma$ is the material conductivity, and $\lambda$ is the spin diffusion length. 
    The subscript ``SIO'' has been omitted for clear representation of the formula. 
    Using the material parameters of SIO listed in Table \ref{tab:SIO}, we calculated the value of $G_\mathrm{SHE}$ and listed in Table \ref{tab:element} 

\subsection{Magnon transport module}\label{subsec:Magnon}
    Pure spin conduction is realized by magnon transport in the insulating ferromagnetic and antiferromagnetic materials, without any charge flow \cite{Behin-Aein2011, Sayed2016}. 
    Following \cite{Sayed2016}, we model the $z$-direction spin transport in BFO with a $\Pi$-network, in which the serial conductor $G_{se}$ represents the spin transmission through the material, and the shunt conductors $G_{sf}$ represent the loss of spins due to spin relaxation. 
    Adopting the concept of tensor spin conductance \cite{Brataas2006, Manipatruni2012} and generalizing Ohm's law to include the spin scattering, we can calculate these two conductance matrices as follows:
    \begin{equation}
    G_{\mathrm{se}}
    =
    \begin{bmatrix}
    0 & 0 & 0 & 0 \\
    0 & 0 & 0 & 0 \\
    0 &  0 & 0  & 0  \\
    0 & 0 &  0 & \frac{\sigma^Swl}{\lambda}\mathrm{csch}\left( \frac{t}{\lambda}\right) \\
    \end{bmatrix},
    \label{eq:Gse}
    \end{equation}
    
    \begin{equation}
    G_{\mathrm{sf}}
    =
    \begin{bmatrix}
    0 & 0 & 0 & 0 \\
    0 & 0 & 0 & 0 \\
    0 & 0 & 0 & 0  \\
    0 & 0 & 0 & \frac{\sigma^Swl}{\lambda}\mathrm{tanh}\left( \frac{t}{2\lambda}\right) \\
    \end{bmatrix},
    \label{eq:Gsf}
    \end{equation}
    where the total spin voltage (current) is a $4\times1$ vector containing the charge voltage (current) and vector spin voltage (current)   $\overline{V} = \left[ V^c \quad V^{s}_x \quad V^{s}_y \quad V^{s}_z\right]$ $\left( \overline{I} = \left[ I^c \quad I^{s}_x \quad I^{s}_y \quad I^{s}_z\right] \right)$.
    The variable $\sigma^S$ stands for the spin conductivity, which indicates the ability of the material to carry magnons along the interested direction; $\lambda$ stands for the spin diffusion length, which is the magnon relaxation length for insulating (anti)ferromagnetic materials. 
    To reiterate, the charge components in Eq. \ref{eq:Gse} and \ref{eq:Gsf} are zero in our case, as BFO possesses excellent electric insulation.
    Since there is only one nonzero component in $G_{se}$ and $G_{sf}$, we simplify their notation to scalars as in:
    \begin{equation}
    G_{\mathrm{se}} = \frac{\sigma^Swl}{\lambda}\mathrm{csch}\left( \frac{t}{\lambda}\right) 
    \label{eq:Gse2}
    \end{equation}
    \begin{equation}
    G_{\mathrm{sf}} = \frac{\sigma^Swl}{\lambda}\mathrm{tanh}\left( \frac{t}{2\lambda}\right) 
    \label{eq:Gsf2}
    \end{equation}
    Substituting the BFO properties listed in Table \ref{tab:SIO} into Eq. \ref{eq:Gse2} and \ref{eq:Gsf2} yields the $G_{se}$ and $G_{sf}$ values shown in Table \ref{tab:element}.

    \subsection{ISHE module}\label{subsec:ISHE}
    The inverse spin Hall effect (ISHE) takes place in the bottom SIO layer, with the same geometry as the top layer. 
    Therefore, we reuse Fig.~\ref{fig:SHmod} for studying ISHE, with emphasis on the distinction betwen SHE and ISHE. 
    In the latter case, we inject a spin current $I^{s}$ from the top surface and measure the charge current $I^c_\mathrm{ISHE}$ flowing out of the material through the right surface.  
    The current-controlled current source of the ISHE module is defined as
    $I^c_\text{ISHE} = \eta_\text{ISHE} I^\text{s}$, where the factor $\eta_\text{ISHE}$ is defined as \cite{Hong2016}
    \begin{equation}
    \eta_{\scaleto{\mathrm{ISHE}}{4pt}} = \theta_\text{ISHE} \frac{\lambda}{l} \mathrm{tanh \frac{t}{2 \lambda}}
    \label{eq:etaISH}
    \end{equation}
    The internal conductance of the current source is obtained by dividing the short-circuit current by the open-circuit voltage, given by
    \begin{equation}
    G_\mathrm{ISHE} = \sigma \frac{w \lambda}{l} \mathrm{tanh} \frac{t}{2 \lambda} \left ( 2 \theta^2_\mathrm{SHE} + \frac{t}{\lambda} \mathrm{coth \frac{t}{2 \lambda}} \right)
    \label{eq:GISH}
    \end{equation}

\begin{table}
\centering
\begin{tabular}{c c c c}\toprule

         Symbols & Parameters & Values & Units \\
         \hline
         $\theta_\mathrm{SHE}$ & spin Hall angle of SIO & 1.00 & 1\\
         $\sigma_{\scaleto{\mathrm{SIO}}{4pt}}$ & electric conductivity of SIO & $1\times 10^5$ & $\mathrm{Siemens} / \mathrm{m}$\\
         $\lambda_{\scaleto{\mathrm{SIO}}{4pt}}$ & spin diffusion length of SIO & 1.4 & nm \\
         \hline
         $\epsilon_{r,\mathrm{BFO}}$ & maximum  relative permittivity of BFO & 2500 & 1 \\
         $\sigma_{\scaleto{\mathrm{BFO}}{4pt}}$ & spin conductivity of BFO $^*$ & $4\times 10^5$ & $\mathrm{Siemens} / \mathrm{m}$\\
         $\lambda_{\scaleto{\mathrm{BFO}}{4pt}}$ & spin diffusion length of BFO $^*$ & 10 & nm \\
         \hline
         $l_{\scaleto{\mathrm{SIO,BFO}}{4pt}}$ & length of SIO (BFO) layer & 50 & $\mathrm{\mu m} $\\
         $w_{\scaleto{\mathrm{SIO,BFO}}{4pt}}$ & width of SIO (BFO) layer & 20 & $\mathrm{\mu m} $ \\
         $t_{\scaleto{\mathrm{SIO,BFO}}{4pt}}$ & thickness of SIO (BFO) layer & 10 & nm \\\bottomrule
\end{tabular}
\caption{Material properties and geometrical dimensions.}
    * Values of spin conductivity and spin diffusion length of BFO are roughly estimated from those of YIG \cite{Cornelissen2015}.
\label{tab:SIO}
\end{table}

    \begin{table}
    \centering
    \begin{tabular}{c c c c}
        \toprule
         Symbols & Parameters & Values & Units \\
         \hline
         $G_\mathrm{IC}$ & metal interconnect conductance & $23.48$ & Siemens \\
         $G_\mathrm{SIO}$ & SIO electrical conductance in horizontal direction & $4.0 \times 10^{-4}$ & Siemens \\
         $G_\mathrm{SIO}^{'}$ & SIO electrical conductance in vertical direction & $1.0 \times 10^{4}$ & Siemens \\
         \hline
         $G_\mathrm{SHE}$ & spin Hall current source internal conductance & $8.0173\times 10^4$ & Siemens \\
         $\eta_\mathrm{SHE}$ & charge-to-spin current conversion rate &  $4.3790 \times 10^3$ & 1\\
         \hline
          $G_\mathrm{se}$ & BFO spin conductance of the serial branch in the $\Pi$ network $^{**}$ & $4.0\times 10^4$ & Siemens \\
          $G_\mathrm{sf}$ & BFO spin conductance of the parallel branch in the $\Pi$ network $^{**}$ & 0.02 & Siemens \\
         \hline
         $G_\mathrm{ISHE}$ & inverse spin Hall current source internal conductance & $5.1182\times 10^{-4}$ & Siemens \\
         $\eta_\mathrm{ISHE}$ & spin-to-charge current conversion rate &  $2.7956 \times 10^{-5}$ & 1\\
         \hline
         $G_\mathrm{AFM-NM(\vv m),33}$ & spin mixing conductance at BFO-SIO interface $^{***}$ &  $3.08 \times 10^{4}$ & Simens\\
         \bottomrule
    \end{tabular}
     \caption{Circuit component values.\\
     ** the spin conductance of BFO is roughly estimated using the YIG properties \cite{Cornelissen2015} \\
     *** the mixed spin conductance value is estimated from \cite{Brataas2006}}
    \label{tab:element}
    \end{table}

\subsection{Spin conductance at ME-SOC interfaces}
One can notice the conductance that connects the SOC layer and the middle ME layer in Fig.~\ref{fig:circuit1}, denoted as $G_\mathrm{AFM-NM}$, with AMF stands for antiferromagnetic (which is our ME material, BFO) and NM stands for nonmagnetic (which is our SOC material, SIO).  
To calculate $G_\mathrm{AFM-NM}$, let us start from the general case of having all three components of the spin current, and tailor the model for the chiral magnon mediated spin orbit logic device. 
    
Consider that the canted magnetic moment $\vv m$ in BFO is oriented in the $\left( \phi =  -45 ^ \circ, \theta = 35.3 ^\circ \right)$ direction, giving that
    $\hat m = sin(\theta)cos(\phi) \hat x + sin(\theta)sin(\phi) \hat y + cos(\theta) \hat z $.
The interface tensor spin conductance is given by \cite{Behin-Aein2011}:
\begin{equation}
        G_\mathrm{AFM-NM(\vv m)} = R(\hat m)^{-1}G_\mathrm{AFM-NM(\vv x)} R(\hat m)
        \label{eq:G_rot}
    \end{equation}
where $R(\hat m)$ is the rotation matrix in terms of the $x$ direction, given by:
\begin{equation}
    R(\hat m)
    =
    \begin{bmatrix}
    1 & 0 \\
    0 & r(\hat m) \\
    \end{bmatrix}, 
    r(\hat m) = 
    \begin{bmatrix}
    \hat m \\
    - \frac{\left( \hat m \times \hat x \right)}{\left || \hat m \times \hat x \right ||}\\
    - \hat m \times \frac{\left( \hat m \times \hat x \right)}{\left || \hat m \times \hat x \right ||}\\
    \end{bmatrix}, 
    \label{eq:R}
\end{equation}
and $G_\mathrm{AFM-NM(\vv x)}$ is the interface tensor spin conductance with $\vv m$ along the $x$ direction, given by:
    \begin{equation}
    \begin{bmatrix}
    I^c\\
    I^{s_x}_x \\
    I^{s_x}_y \\
    I^{s_x}_z \\
    \end{bmatrix}
    =
    \begin{bmatrix}
    0 & 0 & 0 & 0 \\
    0 & G^{\uparrow \uparrow} + G^{\downarrow \downarrow} & 0 & 0 \\
    0 & 0 & 2\mathrm{Re} G^{\uparrow \downarrow} & 2\mathrm{Im} G^{\uparrow \downarrow}  \\
    0 & 0 & -2\mathrm{Im} G^{\uparrow \downarrow} & 2\mathrm{Re} G^{\uparrow \downarrow}  \\
    \end{bmatrix}
    \label{eq:AFM-NM}
    \end{equation}
    where the spin-dependent conductance matrices are given by
    \begin{equation}
    G^{\alpha \alpha} = \frac{e^2}{h} \sum_{nm} \left | t_\alpha ^{nm} \right | ^2 \hspace{0.6cm} \mathrm{ with } \hspace{0.2cm} \alpha = \uparrow \mathrm{or} \downarrow
    \label{eq:Gupup}
    \end{equation}
    and the spin-mixing conductance is given by
    \begin{equation}
    G^{\uparrow \downarrow} = \frac{e^2}{h} \left [ M-\sum_{nm} r_\uparrow ^{nm} \left( r_\downarrow ^ {nm} \right) ^* \right] .
    \label{eq:Gmix}
    \end{equation}
    In both Equation (\ref{eq:Gupup}) and (\ref{eq:Gmix}), $r_{\uparrow (\downarrow)} ^{nm}$ is the reflection coefficient of the up (down) spins for transverse mode $m$ at the SIO side and reflected to transverse mode $n$, and $t_{\uparrow (\downarrow)} ^{nm}$ is transmission coefficient for the up (down) spins approaching from the SIO in transverse mode $m$ and leaving in transverse mode $n$ \cite{Brataas2006}. 
    Since $r_{\uparrow (\downarrow)} ^{nm}$ is usually zero, and we ignore the spin-generated charge current in the compact multiferroic logic, Equation \ref{eq:AFM-NM} reduces to
    \begin{equation}
    \begin{bmatrix}
    I^{s_x}_x \\
    I^{s_x}_y \\
    I^{s_x}_z \\
    \end{bmatrix}
    =
    \begin{bmatrix}
     \frac{\sigma^s wl}{t} & 0 & 0 \\
     0 & \frac{2e^2}{h} \frac{wl k_f^2}{4\pi} & 0  \\
     0 & 0 & \frac{2e^2}{h} \frac{wl k_f^2}{4\pi}  \\
    \end{bmatrix}
    \begin{bmatrix}
    V^{s_x}_x \\
    V^{s_x}_y \\
    V^{s_x}_z \\
    \end{bmatrix}
    \label{eq:AFM-NM_reduced}
    \end{equation}
    and the AFM-NM interface conductance matrix is defined as 
    \begin{equation}
    G_{\mathrm{AFM-NM}}
    =
    \begin{bmatrix}
    G & 0 & 0 \\
    0 & G_{SL}  & 0  \\
    0 & 0 & G_{SL}   \\
    \end{bmatrix}
    =
    \begin{bmatrix}
     \frac{\sigma^s wl}{t} & 0 & 0 \\
     0 & \frac{2e^2}{h} \frac{wl k_f^2}{4\pi} & 0  \\
     0 & 0 & \frac{2e^2}{h} \frac{wl k_f^2}{4\pi}  \\
    \end{bmatrix}
    \label{eq:GAFM-NM}
    \end{equation}
Substituting all the values into Equation (\ref{eq:G_rot}) yields the $G_\mathrm{AFM-NM(\vv m)}$ values shown in Table \ref{tab:element}.
Note that since the focus of this work is on the $z$-direction magnon transport, Table \ref{tab:element} listed only the $G_\mathrm{AFM-NM(\vv m),33}$ component.
    
\section{Compact modeling of chiral magnon mediated spin orbit logic device}
\begin{figure*}
    \centering
    \includegraphics[width=0.6\linewidth]{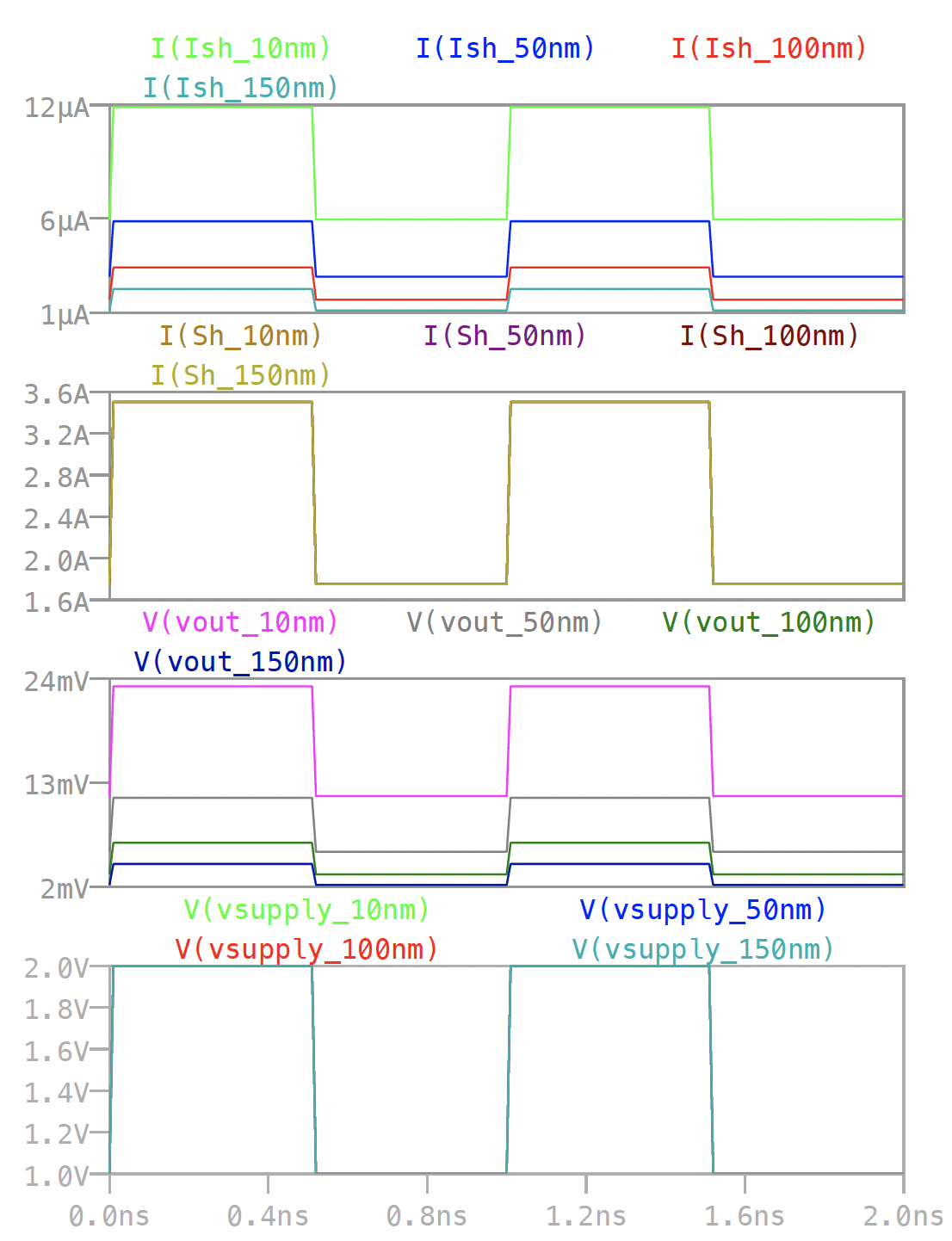}
    \caption{Lossy Buffer logic simulation result for the equivalent circuit in Fig.~\ref{fig:circuit1}(b) where the output follows the input pulse signal characteristics. }
    \label{fig:simulation}
\end{figure*}

In this section, we show preliminary circuit simulation results to demonstrate the spin conduction in BFO affected by its thickness, which is the magnon transport length. 
Fig.~\ref{fig:simulation} shows the output voltage values generated by the same amount of input voltage, with varying BFO thickness $t_\mathrm{BFO}$.
Then $t_\mathrm{BFO}$ increases, the serial BFO spin conductance $G_{se}$ decreases, while the shunt spin conductance $G_{sf}$ increases, indicating degraded magnon transport performance.
Therefore, as $t_\mathrm{BFO}$ changes from 10 nm to 150 nm, the output voltage decreases from 24 mV to 4 mV, which matched our expectation. 
Future works include incorporating the ferroelectric switch process into the simulation, to capture the switch delay time and on/off ratios, as well as optimizing the structure and material design for larger output voltages. 

\newpage
\bibliographystyle{naturemag}
\bibliography{bib_supp}